\documentclass[useAMS, usenatbib, usegraphicx]{mn2e} 

\title[Kinetic AGN feedback with SPH] 
{Kinetic AGN feedback effects on cluster cool cores simulated using SPH} 

\author[P. Barai et al.] 
{Paramita Barai$^{1, 7}$, 
Giuseppe Murante$^{1}$, 
Stefano Borgani$^{1, 2, 3}$, 
Massimo Gaspari$^{4, 5}$, 
\newauthor 
Gian Luigi Granato$^{1}$, 
Pierluigi Monaco$^{1, 2}$, 
Cinthia Ragone-Figueroa$^{6, 1}$ 
\vspace{0.2cm} \\ 
$^{1}$ INAF - Osservatorio Astronomico di Trieste, Via G.B. Tiepolo 11, I-34143 Trieste, Italy \\ 
$^{2}$ Dipartimento di Fisica dell'Universit\`{a} di Trieste, 
Sezione di Astronomia, Via Tiepolo 11, I-34131 Trieste, Italy \\ 
$^{3}$ INFN / National Institute for Nuclear Physics, Via Valerio 2, I-34127 Trieste, Italy \\ 
$^{4}$ Department of Astrophysical Sciences, Princeton University, Princeton, NJ 08544, USA ~~~ 
$^{5}$ Einstein and Spitzer Fellow \\ 
$^{6}$ Instituto de Astronom\'ia Te\'orica y Experimental (IATE),\\ 
Consejo Nacional de Investigaciones Cient\'ificas y T\'ecnicas de la Rep\'ublica Argentina (CONICET),\\ 
Observatorio  Astron\'omico, Universidad Nacional de C\'ordoba, Laprida 854, X5000BGR, C\'ordoba, Argentina\\ 
$^{7}$ current address: Scuola Normale Superiore, Piazza dei Cavalieri 7, I-56126 Pisa, Italy 
(E-mail: paramita.barai@sns.it) \\ 
} 

\begin{document} 

\maketitle 

\label{firstpage} 

\begin{abstract} 

We implement novel numerical models of AGN feedback in the SPH code {\sc GADGET-3}, 
where the energy from a supermassive black hole (BH) is coupled to the surrounding gas in the {\it kinetic} form. 
Gas particles lying inside a bi-conical volume around the BH are 
imparted a one-time velocity ($10,000$ km/s) increment. 
We perform hydrodynamical simulations of isolated cluster (total mass $10^{14} h^{-1} M_{\odot}$), 
which is initially evolved to form a dense cool core, having central $T \leq 10^{6}$ K. 
A BH resides at the cluster center, and ejects energy. 
The feedback-driven fast wind undergoes shock with the slower-moving gas, 
which causes the imparted kinetic energy to be thermalized. 
Bipolar bubble-like outflows form propagating radially outward to a distance of a few $100$ kpc. 
The radial profiles of median gas properties are influenced by BH feedback in the inner regions ($r < 20 - 50$ kpc). 
BH kinetic feedback, with a large value of the feedback efficiency, 
depletes the inner cool gas and reduces the hot gas content, such that 
the initial cool core of the cluster is heated up within a time $1.9$ Gyr, 
whereby the core median temperature rises to above $10^{7}$ K, and the central entropy flattens. 
Our implementation of BH thermal feedback (using the same efficiency as kinetic), 
within the star-formation model, cannot do this heating, where the cool core remains. 
The inclusion of cold gas accretion in the simulations 
produces naturally a duty cycle of the AGN with a periodicity of $100$ Myr.

\end{abstract} 


%
%
%
%

\section{Introduction} 
\label{sec-intro} 

The enormous amounts of energy emitted by the centers of active galaxies are believed to be from 
accretion of matter onto supermassive black holes (SMBHs) lying there 
\citep[e.g.,][]{rees84, Kormendy95, Ferrarese05}. 
Feedback from active galactic nuclei (AGN) strongly influences the formation and 
cosmological evolution of structures, 
whereby the overall properties of a galaxy can be regulated by its central SMBH, 
which also impacts the environment from pc to Mpc scales 
\citep[e.g.,][]{lynden-Bell69, richstone98, King03, Granato04, sazonov05, Barai06, 
Croton06, Barai08, Fabian12, Wagner13}. 
The energy output is often observed as AGN outflows in a wide variety of forms 
\citep[see][for reviews]{Crenshaw03, Everett07}, e.g.: 
collimated relativistic jets and/or overpressured cocoons in radio \citep{Nesvadba08}, 
blue-shifted broad absorption lines in the ultraviolet (UV) and optical \citep{Reichard03, Rupke11}, 
warm absorbers \citep{chartas03, Krongold07} and ultra-fast outflows 
in X-rays \citep{Tombesi13, Feruglio15}, 
molecular gas in far-IR \citep{Feruglio10, Sturm11, Cicone14}. 

A major baryonic component pervading galaxy clusters is the diffuse hot 
(temperature $\sim 10^7 - 10^8$ K) gas present in between galaxies, called the intracluster medium (ICM). 
It emits copiously in the X-ray band, 
as it undergoes radiative cooling by losing energy through the bremsstrahlung process, 
and emission lines of $\sim$keV energy associated to atomic transitions from metals. 
Observations of X-ray radial profiles of the so-called classical {\it cool-core clusters} 
show a sharp peak in the X-ray surface brightness at the core, which imply a low entropy, high density, 
hence a shorter cooling time of the central ICM. 
As the cluster core cools faster, the central gas pressure also decreases. 
The overpressured ICM from the surrounding cluster region should inflow toward the core, 
as a slow, subsonic {\it cooling flow} \citep{Fabian94} in the standard scenario; 
however observations find a lack of emission from cool gas in the cluster center \citep[e.g.,][]{Tamura01}. 


The core cooling times in $\sim 50 \%$ of the clusters is much shorter than the 
cluster lifetimes, few Gyrs \citep{Bauer05, Cavagnolo09}. 
Therefore the dense ICM at cluster cores should cool and fragment to form 
self-gravitating objects like molecular clouds and stars. 
UV observations reveal recent star formation in the central brightest cluster galaxies 
of many cool-core clusters \citep{ODea04, Hicks10}. 
However, enough cold condensed material is not observed, which is called the {\it cooling flow problem}. 
The observed mass deposition rates of cooling and star-forming gas in the cores 
of galaxy clusters with strong cooling flows is much lower than that 
predicted from X-ray observed gas density profiles and simple radiative cooling models 
\citep{Bohringer02, Peterson06, ODea08}. 

The above observations imply that the cooling flow ICM at cluster cores is being heated. 
The most favored mechanism is heating by energy feedback of outflows driven by 
AGN lying in brightest cluster galaxies \citep{Young02, Fabian03, McNamara07, Mittal09, McDonald15}. 
Other possible heating mechanisms are: star-formation and supernovae-driven outflows, 
thermal conduction from the hot outer ICM to the center, cosmic rays, and magnetic fields. 
While X-ray observations reveal a standard pure cooling rate suppression (by less than $10\%$); 
at the same time residual cooling of at least a few percent is observed 
\citep[e.g.][and references therein]{Gaspari15}. 
Hence the heating mechanisms at operation are avoiding overheating of cluster cores. 


The majority ($\sim 70 \%$) of cool-core clusters are observed to host radio sources 
\citep{Burns90, Dunn06}. 
Radio-loud AGN drive relativistic jets on two opposite sides, 
surrounded by overpressured lobes, which expand supersonically and end in bow shocks 
\citep{Blandford76, Kaiser97, Barai07}. 
Radio jets and lobes have complex interactions rising through the cluster atmosphere: 
heat the cooling flow, displace the X-ray emitting gas along the jet direction, 
and/or locally compress the ICM. 
Structures like cavities and bubbles as depressions (or holes) in X-ray surface brightness images 
are observed in many clusters, coincident with the radio jet/lobe emission 
\citep{Carilli94, Bohringer95, McNamara00, Fujita02, Heinz02, Birzan04, Choi04, Reynolds05, Kirkpatrick15}. 

Numerical hydrodynamical simulations are performed to follow the propagation of 
radio jets and lobes injected from the central AGN into the ICM, 
and their subsequent evolution imparting energy and entropy to the gas. 
Several 2D and 3D simulations using Eulerian adaptive-mesh refinement codes have shown that 
mechanical, kinetic or thermal energy coupled from the AGN to surrounding ICM 
is able to offset its radiative cooling. 
Most of these use isolated cluster models, 
an initial isothermal spherically-symmetric analytical profile for the ICM 
\citep{Bruggen02, Basson03, DallaVecchia04, Omma04, Reynolds05-next, AlouaniBibi07, Hillel15}. 
Few studies have included full jet dynamics and a gas feedback model \citep{Vernaleo06, Gaspari12b}, 
used cluster extracted from a cosmological volume \citep{Bruggen05}, 
implemented momentum-driven jets in zoom-in cluster resimulation \citep{Dubois10}. 
Mesh-code idealized simulations demonstrate that strong AGN heating can transform 
a cool-core cluster to a non cool-core state \citep{Guo09}. 

Lagrangian SPH code simulations are also done to study AGN feedback effects on ICM evolution, 
which usually does not consider the radio jet's propagation. 
These SPH studies include additional gas physics; 
star-formation, supernova feedback, and AGN feedback as thermal heating: 
hot buoyant bubbles injected into the ICM during active AGN phases \citep{Sijacki06, Revaz08}; 
transition from quasar-mode to radio-mode feedback when the SMBH accretion rate goes below a given limit, 
and the feedback efficiency is increased by a factor of $4$ \citep{Fabjan10, Hirschmann14}. 
Recently, \citet{Richardson16} compared between adaptive mesh refinement and SPH methods, 
performing cosmological zoom simulations of a galaxy cluster with AGN feedback. 

AGN outbursts are fed by gas accretion from the ICM, so self-regulation may arise naturally. 
When an AGN is active for a period of time, its outflow heating stops the cooling flow. 
This also limits the gas accreting onto the SMBH, turning the AGN dead. 
Afterwards, the ICM cools again, cooling flow restarts, 
providing the central SMBH with fuel and might trigger another phase of AGN outburst. 
The result is a self-regulating cyclic process with cooling flow alternating with bursts of AGN activity. 
Idealized simulations have studied how AGN feedback can successfully self-regulate 
the cooling and heating cycles of the ICM at cluster cores 
\citep{Cattaneo07, Bruggen09, Gaspari11a, Li15, Prasad15}. 
Explicitly simulations including AGN jets have revealed the trigger of a circulation of gas, that is crucial 
in stabilizing cool cores and creating self-regulation \citep[e.g.,][]{OmmaBinney04, Brighenti06}. 

The physical mechanisms of gas accreting onto SMBHs and resulting energy feedback is complex, 
with the surrounding gas likely accelerated by thermal pressure and/or radiation pressure. 
The relevant physical scales are orders of magnitude below 
the scales resolved in current galaxy formation simulations. 
Hence AGN accretion and feedback are incorporated 
in the simulations using {\it sub-resolution} numerical prescriptions \citep[e.g.,][]{SDH05}. 
The physics of the gas, on scales unresolved in cosmological simulations, 
is modelled using spatially averaged properties describing the medium on scales that are resolved. 
Galaxy formation simulations have investigated coupling of AGN feedback using both mechanisms: 
thermal \citep[e.g.,][]{DiMatteo05, Sijacki07, Booth09, Gaspari11b}, 
and kinetic \citep[e.g.,][]{Dubois10, Ostriker10, Barai11a, Gaspari12a, Vazza13}. 


Kinetic AGN feedback in SPH simulations has been implemented in a few previous work. 
\citet{Choi12, Choi14} studied radiative- and momentum-based mechanical AGN feedback 
in isolated and merging galaxies using the GADGET code, 
and \citet{Choi15} performed zoom-in cosmological simulations. 
They implemented output of mass and mechanical momentum (in addition to energy output) 
from the BH to the ambient gas, 
by kicking each wind particle along the direction parallel or anti- parallel to its angular momentum. 

More recently, a similar work was presented by \citet{Zubovas15}. 
Our current study was developed completely independent of this; and their aims/setup are quite different from ours. 
They studied feedback from fixed-luminosity AGN affecting turbulent gas spheres, 
using an initial condition of gas distributed from $100$ pc to $1$ kpc. 
In addition to thermal feedback, there is momentum feedback, where each neighbour particle 
receives a fraction of the AGN wind momentum in a direction radially away from the SMBH. 
Their bi-conical model, where the thermal and kinetic feedback energy are injected in a bi-cone, 
produces a far more complex gas structure around the SMBH. 

We investigate, in this paper, different sub-resolution models and implementations 
of AGN feedback in cluster simulations using the SPH technique. 
We implement novel numerical methods to couple the energy from a central SMBH to the surrounding gas, 
{\it kinetic feedback -} where the velocity (or momentum) of the gas is boosted. 
Here we present simulations of isolated clusters. 
Our goal is to explore the evolution of cluster cool cores, the properties of 
BH accretion, growth and feedback, especially its impact on the cooling ICM. 
We have not done any parameter tuning in this paper to match observations; 
but our aim here is to explore the results of the new implementation, 
especially the differences from our thermal feedback. 


Our previous studies \citep[][and references therein]{Ragone13, Planelles14} used only thermal feedback, 
where the feedback efficiency needed to be increased by a factor of $4$ in the radio-mode. 
\citet{Rasia15} (following \citealt{Steinborn15}) classified the power from AGN into two parts: 
radiative and mechanical; simulating both of them with thermal feedback. 
In this paper, we progressively build a model where 
the effect of {\it radiative power of the AGN is simulated with thermal feedback}, 
and the effect of {\it mechanical power of the AGN is simulated with kinetic feedback}. 
Thus we add to and improve the {\it unified model} of AGN feedback (\S\ref{sec-num-BH-Accr-Feed}), 
by integrating our new kinetic prescription with the mechanical outflow model of \citet{Steinborn15}. 
This renders an SPH code (including the full gas physics required for cosmological simulations) 
able to simulate both quasar-mode and radio-mode AGN, 
and generate bubble-like outflows in the radio-mode (c.f. \S\ref{sec-results-Const-Enrg}) in a self-consistent way. 

This paper is organised as follows: 
we describe our numerical code and simulation setup in \S\ref{sec-numerical}, 
present, analyze and discuss our results in 
\S\ref{sec-results-Const-Enrg} and \S\ref{sec-results-BH-Accr-Feedback} 
(two sections for two categories of simulations), 
discuss the caveats of our study in \S\ref{sec-discussion}, 
and in \S\ref{sec-conclusion} we give a summary of our main findings and discuss future work.

\section{Numerical Method} 
\label{sec-numerical} 

We use a modified version of the TreePM (particle mesh) - 
SPH (smoothed particle hydrodynamics) code {\sc GADGET-3} \citep{Springel05}, 
which includes sub-resolution physics as described next. 
The initial cluster models are presented in \S\ref{sec-num-IC}, 
the BH modules including our new kinetic feedback prescription are detailed in 
\S\ref{sec-num-BH-Kin}-\S\ref{sec-num-Implement}, and our simulations are outlined in \S\ref{sec-num-Sim}. 


Besides AGN feedback, the physical processes implemented in our simulations include: 
radiative physics, star-formation (SF), chemical evolution, and are the same as in \citet{Barai13}. 
As an exception in the simulations with no SF, radiative cooling and heating processes 
are followed for a primordial mix of hydrogen and helium with no metals \citep{Tornatore10}, 
by adopting the cooling rates from \citet{Sutherland93}, including the effect of a UV background \citep{Haardt96}. 


SF is implemented following the multiphase effective sub-resolution model by \citet{SH03}. 
Gas particles with density above a limiting threshold, $\rho_{\rm SF} = 0.13$ cm$^{-3}$ 
(units of number density of hydrogen atoms), contain cold and hot phases, and are star-forming. 
Collisionless star particles are spawned from gas particles undergoing SF, 
based on the stochastic scheme by \citet{Katz96}. 
We allow a gas particle to spawn up to four generations of stars. 

Stellar evolution and chemical enrichment are followed for 11 different elements: 
H, He, C, Ca, O, N, Ne, Mg, S, Si, Fe, 
using the chemical evolution model of \citet{Tornatore07}. 
We include a fixed stellar initial mass function according to \citet{Chabrier03}. 
There is no kinetic feedback from supernovae-driven galactic outflows in our simulations. 
Radiative cooling and heating is computed by adding metal-line cooling from \citet{Wiersma09a}, 
accounting for the 11 species, and photoionizing radiation from the cosmic microwave background 
and the \citet{Haardt01} model for the UV/X-ray background are considered.

\subsection{Initial Condition: Isolated Cluster Model} 
\label{sec-num-IC} 

%

\begin{table*} 
\begin{minipage}{0.71 \linewidth} 
\caption{ 
Isolated Cluster Models. 
Column 2: Total mass (dark matter + gas) inside the virial radius. 
Column 3: Gas mass inside the virial radius. 
Column 4: Number of gas particles within simulation volume. 
Column 5: Mass of each gas particle. 
Column 6: Gravitational softening length. 
} 
\label{Table-Galaxies} 
\begin{tabular}{@{}cccccccccccc} 

\hline 

Series & $M_{\rm tot}$ & $M_{\rm gas}$ & $N_{\rm gas}$ & $m_{\rm gas}$ & $L_{\rm soft}$ \\ 

Name & [$M_{\odot}/h$] & [$M_{\odot}/h$] & & [$M_{\odot}/h$] & (kpc/h) \\ 

\hline 

Fiducial (all other runs) & $10^{14}$ & $10^{13}$ & $2159786$ & $1.5 \times 10^7$ & $2$ \\ 

High Resolution (run {\it c10kHR}) & $10^{14}$ & $10^{13}$ & $21580867$ & $1.5 \times 10^6$ & $1$ \\ 

\hline 

\end{tabular} 
\end{minipage} 
\end{table*} 


\begin{figure*} 
\centering 
\includegraphics[width = \linewidth]{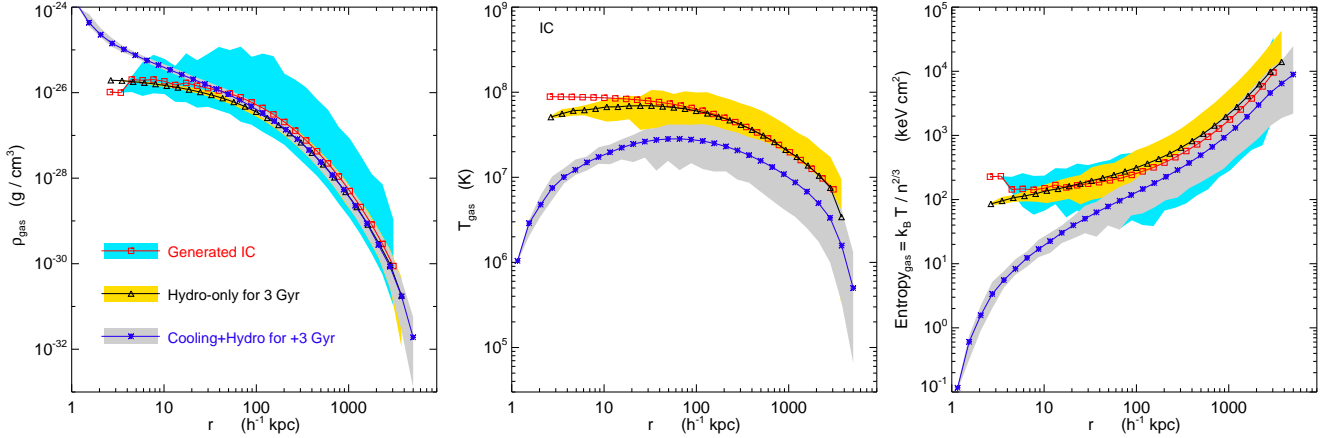} 
\caption{ 
Gas property radial profiles displaying the initial condition processing of the isolated cluster models. 
The plotted curves denote the median quantity in radial bins of each run. 
The respective shaded areas enclose the $90$th percentiles 
above and below the median, showing the radial scatter. 
The red curve (square plotting symbols) with cyan shaded area shows the first generated IC. 
The black curve (triangles) with yellow shaded area is the result from a 
hydrodynamics-only evolution of the first IC up to $3$ Gyr. 
The blue curve (asterisks) with grey shaded area is the result from 
a subsequent cooling+hydrodynamics evolution up to further $3$ Gyr, 
which is taken as the default IC of our simulation runs. 
Left panel is gas density, middle panel is temperature, and right panel is entropy. 
} 
\label{fig-IC_Profiles} 
\end{figure*} 

The initial condition (IC) is an isolated cluster in hydrostatic equilibrium \citep[e.g.,][]{Viola08}. 
Our fiducial cluster has a total mass (including dark matter) of $M_{\rm tot} = 10^{14} h^{-1} M_{\odot}$, 
and virial radius of $R_{\rm 200} = 1.12 h^{-1}$ Mpc. 
A Hubble parameter of $H_{0} = 70.3$ km s$^{-1}$ Mpc$^{-1}$ or $h = 0.703$, 
and baryon density $\Omega_{B, 0} = 0.0463$ are adopted \citep[e.g.][]{Komatsu11}. 
Table~\ref{Table-Galaxies} lists the relevant gas particle masses and force resolutions. 
The baryon fraction of the cluster is $15\%$. 
We sample a sphere of more than $4$ times larger than the virial radius, 
in order to avoid void boundary conditions in the gas just at the virial radius, 
which makes the total gas mass to be $3.24 \times 10^{13} h^{-1} M_{\odot}$. 
There are no stars in the IC. 
In the runs including SF, new star particles form during the simulation via star formation in the gas. 

All the particles (gas, stars, BH) follow collisionless gravitational dynamics, 
while in addition the gas particles undergo hydrodynamical interactions. 
The Plummer-equivalent softening length for gravitational forces is set to 
$L_{\rm soft} = 2 h^{-1}$ kpc for all the particles, in our default resolution simulations. 
The SPH computations are carried out using a $B$-spline kernel \citep{Monaghan85}, 
which is the standard one used in the {\sc GADGET} code (Eq.~4 of \citealt{Springel05}). 
The number of neighbors for each gas particle is taken as $64$. 
We set a minimum value of the SPH smoothing length to be $0.1$ times the gravitational softening. 
Note that, using isolated ICs, we could have reached much larger force and mass resolution. 
Instead, our choice here is to mimic the typical resolutions of 
cosmological, zoomed-in simulations of galaxy clusters. 

The procedure of generating our IC consists of the following steps, which are described in detail in the next paragraphs. 
(a) Density profiles for dark matter and hydrostatic gas are computed analytically. 
(b) The gas density profile is sampled to produce the initial cluster. 
(c) The cluster is evolved with gravity and hydrodynamics, for several dynamical times, 
to reach true equilibrium and relaxation. 
(d) The cluster is then evolved with gravity, hydrodynamics and cooling, to develop a cool core. 

The cluster dark matter halo is represented by a background fixed static potential, 
following the \citet{Navarro97} (NFW) density profile, 
\begin{equation} 
\label{eq-NFW} 
\rho(r) = \rho_{\rm crit} \frac{\delta_c}{\left( r/r_s \right) \left( 1 + r/r_s \right)^2}, 
\end{equation} 
where $r_s$ is a scale radius, $\delta_c$ is a characteristic (dimensionless) density, 
and $\rho_{\rm crit}$ is the critical cosmic density. 
Some studies \citep[e.g.][]{Ragone12} find an {\it adiabatic de-contraction} effect 
on the dark matter halo due to massive outflows generated by AGN. 
Nevertheless, we ignore any impact of AGN on the dark matter halo 
(by assuming it as a static potential); 
because the goal of our idealized simulations is to study the 
different forms of feedback effects on the cluster-central gas. 

A spherical distribution of gas denotes the ICM, initially in hydrostatic equilibrium within the NFW potential. 
The equilibrium solution for the gas can be found \citep{Suto98}, 
assuming that the baryonic fraction of the halo is negligible and the gas 
(of density $\rho_{\rm g}$, and pressure $P_{\rm g}$) 
follows a polytropic equation of state with index $\gamma_{\rm p}$: 
$P_{\rm g} \propto \rho_{\rm g}^{\gamma_{\rm p}}$. 
The equation of hydrostatic equilibrium is: 
\begin{equation} 
\label{eq-hydrostatic} 
\frac{dP_{\rm g}}{dr} = -G \frac{\rho_{\rm g} M(< r)}{r^2}. 
\end{equation} 
Here $M(< r)$ is the halo mass within radius $r$, which can be derived from the NFW profile. 
With these assumptions, Eq.~(\ref{eq-hydrostatic}) can be solved analytically \citep{Komatsu01}, 
as done by \citet{Viola08}, which we follow. 
The analytic profiles for dark matter and hydrostatic gas are thus computed, 
and sampled to produce the initial gas distribution. 



This initially generated isolated cluster is processed to form a cool-core, 
as displayed in Fig.~\ref{fig-IC_Profiles} showing the radial profiles of 
gas density (left panel), temperature (middle), and entropy (right). 
The first generated IC is the red curve with square plotting symbols; the cyan shaded area 
reveals the scatter (most prominent in the density profile) due to numerical noise. 
The IC is first allowed to evolve using hydrodynamical interactions only up to $3$ Gyr 
(corresponding to a few dynamical times for such a cluster), 
the result of which is shown by the black curve (triangles) with yellow shaded area. 
This process relaxes the IC while getting rid of the initial noise, and reaches an equilibrium. 
The initial noise decays rather quickly during the hydrodynamics-only evolution, 
within a time of $0.05$ Gyr, which is about $5\%$ of the dynamical time of the cluster. 


The hydro-only result is subsequently evolved using cooling (primordial composition gas with no metals) 
and hydrodynamics, up to further $3$ Gyr, which produces the blue curve (asterisks) with grey shaded area. 
The cooling evolution forms a cold, dense core at the cluster center. 
The cluster at this stage (blue asterisks) is taken as the default IC of our simulation runs. 
The mass cooling rate (to temperature $\leq 5 \times 10^{5}$ K, within $100$ kpc of the cluster) 
in our IC is $6 M_{\odot}$/yr. 


After this, rescaled as $t = 0$, one BH is created in our simulations residing at the center of cluster, at rest. 
It is a collisionless particle in the {\sc GADGET-3} code, having a 
{\it dynamical} mass ($m_{\rm BH,dyn}$) equal to a gas particle mass, given by simulation resolution. 
The BH outputs feedback energy: a fixed power in the first cases (\S\ref{sec-num-BH-Const-Enrg}), 
and next undergoing mass growth and resulting feedback (\S\ref{sec-num-BH-Accr-Feed}).

\subsection{Kinetic AGN Feedback} 
\label{sec-num-BH-Kin} 

Feedback from BH corresponds to an output power $\dot{E}_{\rm feed}$ 
injected and coupled to the surrounding gas. 
It drives gas outward at a velocity $v_w$ and mass outflow rate $\dot{M}_w$. 
The energy-conservation equation, with the kinetic energy of the outflowing gas 
equated to the feedback energy from the BH, is, 
\begin{equation} 
\label{eq-Energy-Conservation} 
\frac{1}{2} \dot{M}_w v_w^2 = \dot{E}_{\rm feed} . 
\end{equation} 
This is similar to the {\it energy-driven wind} formalism of \citet{Barai14}. 
Typical AGN wind velocity values seen in observations is between 
$v_w = {\rm few} ~ 1000 - 10000$ km/s \citep[e.g.,][]{Perna15}. 
We consider $v_w$ as a free model parameter. 
This is a simplified assumption of our model, 
which is intended to be applied to cosmological simulations, 
although more physically the AGN wind velocity should be self-regulated \citep[e.g.,][]{Gaspari11b}. 


\subsection{First Case: Constant BH Energy} 
\label{sec-num-BH-Const-Enrg} 

In the first set of runs, the central BH is considered to output a fixed energy 
(BH accretion is not taken into account here), 
i.e. $\dot{E}_{\rm feed}$ = constant, at a fixed duty-cycle. 
These runs have only radiative cooling, and no SF, 
in order to easily isolate the effect of AGN energy injection on the gas. 

The default BH power value that we explore is $\dot{E}_{\rm feed} = 10^{45}$ erg/s; 
with two runs having higher and lower powers. 
The BH activity is considered to be on for $50$ Myr when the $\dot{E}_{\rm feed}$ energy is ejected, 
and then off for $100$ Myr when there is no energy output. 
Such a on/off fixed-power duty-cycle occurs periodically up to the end of the simulation, for a few Gyrs. 







\subsection{Second Case: BH Accretion and Energy Feedback} 
\label{sec-num-BH-Accr-Feed} 

The second set of runs includes star-formation, 
and a BH undergoing gas accretion, mass growth and resulting feedback. 
The accretion of surrounding gas onto a SMBH of mass $M_{\rm BH}$ is 
parametrized by the Bondi-Hoyle-Lyttleton rate \citep{Hoyle39, Bondi44, Bondi52}: 
\begin{equation} 
\label{eq-Mdot-Bondi} 
\dot{M}_{\rm Bondi} = \alpha \frac{4 \pi G^2 M_{\rm BH}^2 \rho}{ \left(c_{s}^2 + v^2\right) ^ {3/2}}. 
\end{equation} 
Here, $\rho$ is the gas density, $c_{s}$ is the sound speed, and $v$ is the velocity of the BH relative to the gas. 
The Bondi radius ($= G M_{\rm BH} / c_{s}^2$) 
and sonic point (where the gas passes through Mach number $= 1$, 
going from subsonic velocities at larger radii to supersonic at smaller radii) 
are of the order of $\sim 10$'s of pc, hence are unresolved in current galaxy formation simulations. 
The gas properties ($\rho, c_{s}$) are estimated by smoothing 
on the resolution scale (kpc to few $100$'s of pc) at the BH location. 
This results in artificially low densities compared to spatially resolving the Bondi radius scale. 
Furthermore, smaller-scale simulations \citep{Barai12, Gaspari13, Gaspari15b} 
show that the cooling gas is multiphase, with a variable accretion rate. 
This cold phase of the ISM is not resolved in galaxy simulations. 

As a numerical correction, the accretion rate is enhanced in the simulations 
by setting the multiplicative factor to $\alpha = 100$ 
\citep[e.g.,][]{SDH05, Sijacki07, DiMatteo08, Khalatyan08, Johansson09a, Sijacki09, Dubois13}, which we adopt. 
Most of our simulations are done using a single mode of accretion 
considering all the gas (of any temperature) around BH. 
In one run we employ the cold accretion model by \citet{Steinborn15}, 
using $\alpha_{\rm cold} = 100$ for computing 
the accretion rate of the cold gas (temperature $T \leq 5 \times 10^{5}$ K), 
and $\alpha_{\rm hot} = 0$ for the hot gas ($T > 5 \times 10^{5}$ K). 

Accretion onto the BH is limited to the Eddington rate, making the resultant accretion rate, 
\begin{equation} 
\label{eq-Mdot-BH} 
\dot{M}_{\rm BH} = {\rm min} \left( \dot{M}_{\rm Bondi}, \dot{M}_{\rm Edd} \right). 
\end{equation} 
Here, $\dot{M}_{\rm Edd}$ is the Eddington mass accretion rate, 
expressed in terms of the Eddington luminosity, 
\begin{equation} 
\label{eq-LEdd} 
L_{\rm Edd} = \frac{4 \pi G M_{\rm BH} m_p c} {\sigma_T} = \epsilon_r \dot{M}_{\rm Edd} c^2 , 
\end{equation} 
where, $G$ is the gravitational constant, $m_p$ is the mass of a proton, $c$ is the speed of light, 
and $\sigma_T$ is the Thomson scattering cross-section for an electron. 

A fraction of the accreted rest-mass energy is radiated away by the  BH. The radiation luminosity is, 
\begin{equation} 
\label{eq-Lr-BH} 
L_r = \epsilon_r \dot{M}_{\rm BH} c^2, 
\end{equation} 
with $\epsilon_r$ being the radiative efficiency. 
We adopt the mean value for radiatively efficient accretion onto a Schwarzschild BH 
\citep{Shakura73}: $\epsilon_r = 0.1$, which is held at a fixed value. 
A fraction $\epsilon_f$ of this radiated energy is coupled 
to the surrounding gas as feedback energy from the BH: 
\begin{equation} 
\label{eq-Edot-Feed} 
\dot{E}_{\rm feed} = \epsilon_f L_r = \epsilon_f \epsilon_r \dot{M}_{\rm BH} c^2. 
\end{equation} 
We consider the feedback efficiency $\epsilon_f$ as a free parameter in our models. 

We examine different ways by which the BH feedback energy can be coupled to the gas: 

\noindent $\bullet$ {\bf Thermal} : 
We adopt the default scheme from \citet{SDH05}. 
The energy $\dot{E}_{\rm feed}$ is distributed thermally to heat up the gas isotropically around the BH. 
The temperature of the neighboring gas particles 
(those contributing to Eq.~\ref{eq-BH-Smooth} in \S\ref{sec-num-Implement}, 
particles lying within the smoothing length of the BH) 
is incremented by an amount scaled by their SPH kernel weights. 



\noindent $\bullet$ {\bf Kinetic} : 
The kinetic energy of the outflowing gas is derived from the BH radiation luminosity which is fed back. 
The energy-conservation equation becomes: 
\begin{equation} 
\frac{1}{2} \dot{M}_w v_w^2 = \dot{E}_{\rm feed} = \epsilon_f \epsilon_r \dot{M}_{\rm BH} c^2 . 
\end{equation} 
This gives the outflow rate in terms of the BH accretion rate, 
\begin{equation} 
\label{eq-MdotW-EDW} 
\dot{M}_w = 2 \epsilon_f \epsilon_r \dot{M}_{\rm BH} \frac{c^2}{v_w^2} . 
\end{equation} 
There are two free parameters in our sub-resolution model of kinetic feedback: $\epsilon_f$ and $v_w$. 
These parameters can be varied to obtain a closest match of the simulation versus 
observational $[M_{\rm BH} - \sigma_{\star}]$ relation \citep{Barai14}. 
The best-fit $\epsilon_f$ and $v_w$ however depends strongly 
on the simulation characteristics (isolated galaxy/cluster, galaxy merger, cosmological box, etc). 
We decide not to attempt for obtaining any best-fit parameters in this isolated cluster work, 
which is a study of how kinetic BH feedback affects cluster cool cores. 

\noindent $\bullet$ {\bf Unified} : 
The unified scheme is a combination of BH thermal and kinetic feedback, at every timestep. 
Following \citet{Steinborn15}, 
this model considers two components of AGN feedback \citep[based on][]{Churazov05}: 
{\it outflow} - mechanical feedback with outflow power $P_o$, 
and {\it radiation} - radiative component with luminosity $L_r$. 
The total feedback energy per unit time is then the sum (their Eqs.~7-9): 
\begin{equation} 
\label{eq-Edot-Unified} 
\dot{E}_{\rm feed} = P_o + \epsilon_f L_r = (\epsilon_o + \epsilon_f \epsilon_r) \dot{M}_{\rm BH} c^2. 
\end{equation} 
\citet{Steinborn15} implemented both the components as thermal feedback. 
Here in our unified model, the mechanical outflow power $P_o$ is coupled to the gas by kinetic feedback.

\subsection{Implementation in the {\sc GADGET-3} code} 
\label{sec-num-Implement} 

Quantities in the local environment around the BH are computed 
in a kernel-weighted way, by smoothing over neighboring gas particles. 
The kernel size, or the BH smoothing length $h_{\rm BH}$, is determined at each timestep 
(in analogy to finding gas particle smoothing length) by implicit solution of the equation, 
\begin{equation} 
\label{eq-BH-Smooth} 
\frac{4}{3} \pi h_{\rm BH}^3 \rho_{\rm BH} = M_{\rm ngb} . 
\end{equation} 
Here $\rho_{\rm BH}$ is the kernel estimate of the gas density at the position of the BH, 
and $M_{\rm ngb}$ is the mass of $\sim 4 \times 64$ neighboring gas particles. 
The BH particle has $4$ times more neighbors \citep[e.g.,][]{Fabjan10} than an SPH gas particle, 
in order to have more particles for computing the physical properties of the BH sub-resolution model. 

We distribute the BH kinetic feedback energy to the gas lying inside a bi-cone around the BH, 
with a motivation to create a bipolar narrow-angled outflows. 
A bi-conical volume is defined with the BH at the apex, 
and two cones on opposite sides of the BH, having their axes along two diametrically opposed directions. 
The slant height of each cone is $h_{\rm BH}$, and the total opening angle of a single cone is taken as $60^{\circ}$. 
The cone-axes directions are considered as fixed in most of our runs, 
and we explore one case of varying this direction (as written in Table~\ref{Table-Sims}). 
The fixed direction is taken along $\pm z$-axis, 
for the BH located at the origin of our coordinate system. 
The gas particles lying within this bi-cone volume around the BH are tracked, 
and their total mass $M_{\rm gas}^{\rm cone}$ is computed. 



The feedback energy is distributed to gas within a distance $h_{\rm BH}$ from the BH. 
In our simulations $h_{\rm BH}$ typically lies between $(0.1 - 5)$ kpc. 
Gas particles inside the bi-cone are stochastically selected and kicked into AGN wind, 
by imparting a one-time $v_w$ velocity boost. 
We use a probabilistic criterion 
(similar to other sub-resolution prescriptions in {\sc GADGET-3}) for the implementation. 
A probability for being kicked is calculated in a timestep $\Delta t$ 
for $i$'th gas particle within bi-cone: 
\begin{equation} 
\label{eq-prob-Kin} 
p_i = \frac{\dot{M}_w \Delta t} {M_{\rm gas}^{\rm cone}} . 
\end{equation} 
Here $\dot{M}_w$ is the mass outflow rate obtained from Eq.~(\ref{eq-Energy-Conservation}). 
At a given timestep, all the gas particles within the bi-cone have the same probability to be ejected. 
A random number $x_i$, uniformly distributed in the interval $[0, 1]$, is drawn and compared with $p_i$. 
For $x_i < p_i$, the gas particle is given a wind velocity kick. 

The inverse proportionality of $p_i$ with $M_{\rm gas}^{\rm cone}$ ensures that 
the number of particles kicked does not depend on the geometry of the volume, but depends on $\dot{M}_w$ only. 
The quantity $\dot{M}_w \Delta t$ is the mass of gas to be kicked. 
The probability is constructed such that the available particles (total mass $M_{\rm gas}^{\rm cone}$ within bi-cone) 
are sampled to reproduce kicking at the rate given by $\dot{M}_w$, on average. 
During a simulation, we always ensure that $p_i < 1$. 


After receiving AGN wind kick, a gas particle's velocity becomes: 
\begin{equation} 
\label{eq-vNew} 
\vec{v}_{\rm new} = \vec{v}_{\rm old} + v_w \hat{z}, ~~~ {\rm or}, ~~~ 
\vec{v}_{\rm new} = \vec{v}_{\rm old} - v_w \hat{z} . 
\end{equation} 
The kick direction ($+ \hat{z}$ or $- \hat{z}$) is set along the axis of the cone 
inside which the gas particle lies, depending on the location of the particle w.r.t.~the BH. 
Our choice represents a scenario where one BH injects kinetic energy along a fixed direction, 
chosen here as the $z$-axis for numerical simplicity. 
Such a prescription allows the energy injected to be channelized most efficiently, 
creating bi-directional outflows originating from the central BH. 
In future cosmological simulations, the fixed direction for AGN kick would represent the spin of the BH. 

Some studies \citep[e.g.,][]{Barai13} implement hydrodynamic decoupling 
of the gas particles kicked into wind by supernovae, 
to enable the outflow to escape without affecting star-formation in the galaxy. 
Here we perform runs considering both decoupling and coupling of the AGN wind particles 
from hydrodynamic interactions. 


In the presence of BH accretion/growth (\S\ref{sec-num-BH-Accr-Feed}), 
the collisionless BH particle (\S\ref{sec-num-IC}) has a seed BH of 
initial mass $M_{\rm BH} = M_{\rm BH, seed} = 10^5 M_{\odot}$. 
At each timestep $\Delta t$, it grows according to the BH accretion rate (Eq.~\ref{eq-Mdot-BH}), 
its mass increases by an amount $\dot{M}_{\rm BH} \Delta t$, 
with the dynamical mass $m_{\rm BH,dyn}$ (\S\ref{sec-num-IC}) remaining the same. 
After a BH has grown such that $M_{\rm BH} \geq m_{\rm BH,dyn}$, 
it might accrete (or {\it swallow}) neighboring gas particles, using a stochastic methodology. 
When a gas particle is swallowed, it is removed from the simulation, 
and $m_{\rm BH,dyn}$ increases by the swallowed particle mass $m_{\rm gas}$. 
This conserves dynamical mass within the computational volume. 
The probability of swallowing gas is set to ensure that $M_{\rm BH}$ and $m_{\rm BH,dyn}$ 
track each other closely. 

Once the BH is created residing at the cluster center at $t = 0$ (\S\ref{sec-num-IC}), 
it is free to move at subsequent times. 
We have checked that in our simulations, 
the BH remains within a displacement $0.1$ to $0.2$ kpc from the cluster center; 
which means that the BH does not advect away. 




\subsection{Simulations} 
\label{sec-num-Sim} 

%

\begin{table*} 
\begin{minipage}{0.9 \linewidth} 
\caption{ 
Isolated cluster simulation runs and parameters. 
Column 1: Name of simulation run. 
Column 2: AGN accretion mode (relevant for the second-case BH-accretion runs). 
Column 3: Specifications of AGN feedback model. 
Column 4: Output BH power value, in the first-case constant-energy runs. 
Column 5: Feedback efficiency, $\epsilon_f$, in the second-case BH-accretion runs. 
Column 6: $v_w$ = Outflow velocity in kinetic feedback prescription. 
Column 7: Direction along which velocity kick is given. 
Column 8: Coupling or decoupling of wind particles from hydrodynamic interactions. 
} 
\label{Table-Sims} 
\begin{tabular}{@{}cccccccc} 

\hline 

Run & AGN Accretion & AGN Feedback & $\dot{E}_{\rm feed}$ & $\epsilon_f$ & $v_w$ & Direction & Hydrodynamic coupling \\ 
Name & & & [erg/s] & & [km/s] & \\ 

\hline 
\multicolumn{8}{c}{cooling-only, no-SF, constant energy output from BH at fixed duty-cycle} \\ 
\hline 

{\it c5k} & & Constant energy & $10^{45}$ & & $5,000$ & Fixed & Coupled \\  

{\it c10k}, {\it c10kHR} & & Constant energy & $10^{45}$ & & $10,000$ & Fixed & Coupled \\  

{\it c5kDec} & & Constant energy & $10^{45}$ & & $5,000$ & Fixed & Decoupled up to 50kpc \\  

{\it c10kDec} & & Constant energy & $10^{45}$ & & $10,000$ & Fixed & Decoupled up to 50kpc \\  


{\it c10kDecHigh} & & Constant energy & $4 \times 10^{45}$ & & $10,000$ & Fixed & Decoupled up to 50kpc \\  

{\it c10kLow} & & Constant energy & $2 \times 10^{44}$ & & $10,000$ & Fixed & Coupled \\  

{\it c10kRand} & & Constant energy & $10^{45}$ & & $10,000$ & Random & Coupled \\  


\hline 
\multicolumn{8}{c}{with cooling, SF and BH growth, energy output from BH accretion rate} \\ 
\hline 

{\it SF} & & No BH & & & & & \\  

{\it thr} & All gas & Thermal & & $0.02$ & & Isotropic & \\  

{\it kin10k} & All gas & Kinetic & & $0.02$ & $10,000$ & Fixed & Coupled \\  


{\it unf10k} & All gas & Unified & & $0.02$ & $10,000$ & Fixed & Coupled \\  

{\it unf5k} & All gas & Unified & & $0.02$ & $5,000$ & Fixed & Coupled \\  

{\it unf10kLo} & All gas & Unified & & $0.002$ & $10,000$ & Fixed & Coupled \\  

{\it unf10kHi} & All gas & Unified & & $0.2$ & $10,000$ & Fixed & Coupled \\  

{\it unf10kLoCA} & Cold gas & Unified & & $0.002$ & $10,000$ & Fixed & Coupled \\  


\hline 
\end{tabular} 

\end{minipage} 
\end{table*} 


Table~\ref{Table-Sims} lists the series of simulations we perform. 
They form two broad categories, as written below. 
Each category incorporate the same non-AGN sub-resolution physics, 
and the different runs in it investigate various AGN feedback models. 
The evolution is followed for a time $2$ to $2.5$ Gyr in the different runs. 

\begin{itemize} 

\item {\it Cooling only, no SF, constant energy output from BH at fixed duty-cycle, 
coupled as kinetic feedback} 
(\S\ref{sec-num-BH-Const-Enrg}): 
The gas is considered to be of primordial composition, i.e., hydrogen and helium only, with no metals. 
The run names correspond to the variation of the model parameters as written in the following. 

\begin{itemize} 

\item {\it c5k} : outflow velocity $v_w = 5,000$ km/s. 

\item {\it c10k} : $v_w = 10,000$ km/s. 

\item {\it c10kHR} : high-resolution run with $v_w = 10,000$ km/s. 

\item {\it c5kDec}, {\it c10kDec} : different-$v_w$ runs with the wind particles 
decoupled from hydrodynamic interactions up to a distance of $50$ kpc. 


\item {\it c10kDecHigh} : wind-decoupled run, with a higher output BH power 
$\dot{E}_{\rm feed} = 4 \times 10^{45}$ erg/s. 

\item {\it c10kLow} : lower value of output BH power $\dot{E}_{\rm feed} = 2 \times 10^{44}$ erg/s. 

\item {\it c10kRand} : direction of kicking wind particles (bi-cone axis) 
changes randomly between duty cycles, i.e. from one activity cycle of BH to another. 

\end{itemize} 

\item {\it With cooling, SF and BH growth, energy output from BH accretion rate} 
(\S\ref{sec-num-BH-Accr-Feed}): 
Metals are generated according to the stellar evolution and chemical enrichment model 
described in \S\ref{sec-numerical}. 
Six of our simulations have a single mode of accretion considering all the gas (of any temperature) around BH, 
using $\alpha = 100$ in the Bondi accretion rate. 
One run has the cold accretion model \citep{Steinborn15}, using $\alpha_{\rm cold} = 100$ and $\alpha_{\rm hot} = 0$. 

\begin{itemize} 

\item {\it SF} : 
star-formation, stellar evolution, and chemical enrichment only (no BH). 

\item {\it thr} : 
thermal feedback from BH, with $\epsilon_f = 0.02$. 

\item {\it kin10k} : 
kinetic feedback from BH, with $\epsilon_f = 0.02$, $v_w = 10,000$ km/s. 

\item {\it unf$\star$} : 
unified feedback model (a combination of BH thermal and kinetic feedback), 
where the mechanical outflow power is coupled to the gas in the kinetic form. 

\begin{itemize} 

\item {\it unf10k} : $\epsilon_f = 0.02$, $v_w = 10,000$ km/s. 

\item {\it unf5k} : $\epsilon_f = 0.02$, $v_w = 5,000$ km/s. 

\item {\it unf10kLo} : $\epsilon_f = 0.002$, $v_w = 10,000$ km/s. 

\item {\it unf10kHi} : $\epsilon_f = 0.2$, $v_w = 10,000$ km/s. 

\item {\it unf10kLoCA} : 
cold gas accretion model, unified feedback, with $\epsilon_f = 0.002$, and $v_w = 10,000$ km/s. 

\end{itemize} 

\end{itemize} 

\end{itemize} 

Note that the escape velocity of our cluster halo is 
$v_{\rm esc} = \sqrt{2 G M_{\rm tot} / R_{\rm 200}} = 840$ km/s. 
Therefore both the values of $v_w$ explored in our models are above the cluster escape speed.

\section{Results: Constant-Energy Output from BH at Fixed Duty-Cycle} 
\label{sec-results-Const-Enrg}

\subsection{Formation of Bubble-like Outflows} 
\label{sec-res-outflow} 

\begin{figure*} 
\centering 
\includegraphics[width = 0.7 \linewidth]{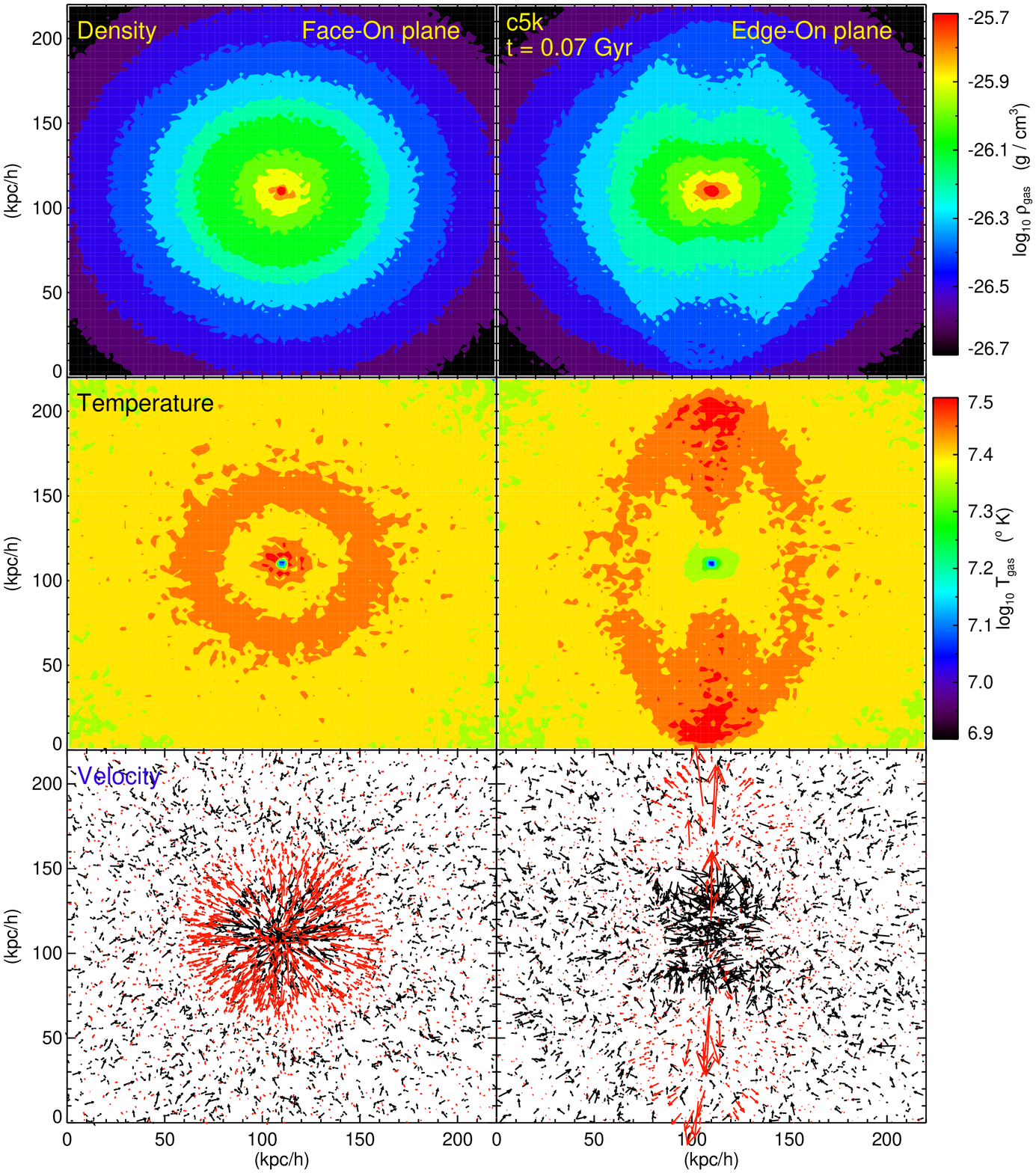} 
\caption{ 
Projection in 2D of gas kinematics in the constant-energy run 
{\it c5k} ($v_w = 5,000$ km/s) at a time $t = 0.07$ Gyr. 
The two columns present the face-on (left) and edge-on (right) planes 
of a $(200 h^{-1}$ kpc$)^3$ volume around the BH at the center of the cluster. 
First row shows density, second row is temperature, 
all projected values, colour coded from red as the highest and black as the lowest. 
Third row depicts the velocity vectors of $10 \%$ of all the gas particles within the projected volume, 
with the outflowing ($v_r > 0$) particles denoted as red, and the inflowing ($v_r < 0$) as black. 
} 
\label{fig-Projected_Map_Run13_snap005} 
\end{figure*} 

Kinetic energy feedback imparted to the gas creates bipolar bubble-like outflows, 
originating from the central BH, and propagating radially outward up to a few $100$ kpc. 
The morphology and structure of a representative outflow for the constant-energy case 
is plotted in Fig.~\ref{fig-Projected_Map_Run13_snap005}, showing the projected gas kinematics 
in the inner $(200 h^{-1}$ kpc$)^3$ region of run {\it c5k} at an evolution time $0.07$ Gyr. 
Gas is kicked with $v_w = 5,000$ km/s in bi-cone volume around the BH along a fixed direction 
(revealed by the red outflowing velocity arrows in the bottom-right panel), and propagates outward. 
It shocks with the surrounding slower-moving gas, and 
leads to the formation of outflowing bubble-like structures. 
The developed outflowing bubbles are hot (visible as red areas in the temperature plot), 
in the face-on plane (middle-left panel) shaped as spherical-shell, and in the edge-on plane 
(middle-right panel) as extended bipolar terminated with an outer elliptical-shaped shock. 
The bubbles consist of low-density gas (first row). 

The time of $0.07$ Gyr plotted in Fig.~\ref{fig-Projected_Map_Run13_snap005} 
corresponds to a part of the first outburst cycle of the BH. 
The dense core of the cluster (red areas in the density panels) is still cool at this time, 
visible as the blue central region in the temperature panels. 
However, the cool-core is destroyed within three or four activity cycles of the BH, 
and the center is heated up, as will be described later. 

\begin{figure*} 
\centering 
\includegraphics[width = 0.8 \linewidth]{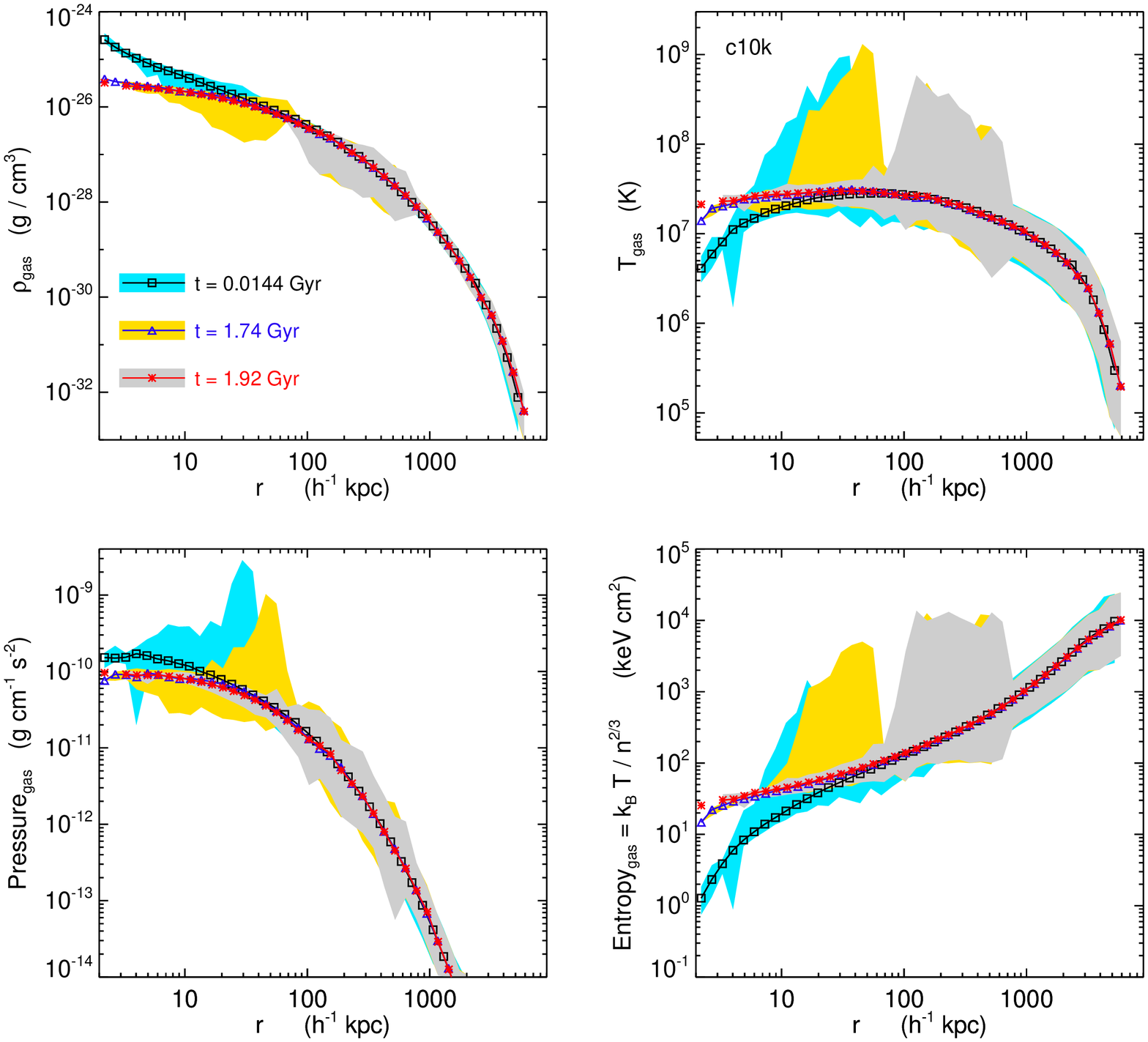} 
\caption{ 
Radial profiles of gas properties at three different epochs in run {\it c10k} ($v_w = 10,000$ km/s): 
density (top-left panel), temperature (top-right), pressure (bottom-left), and entropy (bottom-right panel). 
The curves denote the median quantity (in cgs units) in radial bins, and 
the shaded areas enclose the $90$th percentiles above and below the median, showing the radial scatter. 
The plotted times are: 
$t = 0.01$ Gyr (black curve with square symbols and cyan shaded area) showing the first outflow propagation, 
$t = 1.74$ Gyr (blue curve with triangles and yellow shaded area) a later time AGN-on active epoch, and 
$t = 1.92$ Gyr (red curve with asterisks and grey shaded area) an inactive epoch when the AGN is off. 
} 
\label{fig-GasProperties_vs_R_Run14} 
\end{figure*} 

Fig.~\ref{fig-GasProperties_vs_R_Run14} presents the radial profiles of gas properties 
at three different time epochs in run {\it c10k}. 
The curves denote the median value of the relevant property in radial bins, and 
the shaded areas enclose the $90$th percentiles above and below the median, showing the radial scatter. 
The median quantities at the three different times are indistinguishable from each other at larger radii: $r > 50$ kpc. 
BH feedback imparted to the gas affects the median property at $r \leq 50$ kpc, 
and the radial scatter up to further out $r \sim 700$ kpc. 
The central gas kicked into wind with $v_w = 10,000$ km/s thermalizes its kinetic energy 
in the form of shocks, and generates bubble-like structures propagating outward. 
The outflows are visible as rise in the radial scatter (shaded areas) in Fig.~\ref{fig-GasProperties_vs_R_Run14}: 
hot (top-right panel) reaching temperature as high as $T = 10^{9}$ K, 
high-entropy (bottom-right), low-density (top-left), and high-pressure (bottom-left). 

The earliest time plotted: $t = 0.01$ Gyr (black curve with square symbols) 
is an epoch of the first active period of the BH. 
The cyan shaded area shows the first outflow propagation, extending over $r = 6 - 40$ kpc. 
During every AGN-on active period ($50$ Myr), such a large outflow originates from the center and 
disperse out, moving up to a distance few $100$ kpc during the time the central AGN is off ($100$ Myr); 
before another one starts at the next active cycle. 
The region $r = 100 - 700$ kpc continues to accumulate these hot, high-entropy, low-density bubbles; 
visible as the yellow and grey shaded areas Fig.~\ref{fig-GasProperties_vs_R_Run14}. 
The blue curve with triangles ($t = 1.74$ Gyr) is a later time AGN-on active epoch at the $11$th cycle; 
the yellow shaded area displays both an originating bubble at $10$s kpc, 
and the previous accumulated bubbles at $100$s kpc. 
The red curve with asterisks ($t = 1.92$ Gyr) is an inactive epoch after the $12$th cycle when the AGN is off; 
the grey shaded area displays the previous accumulated bubbles at $100$s kpc. 
The bubble-like outflows propagate radially outward up to $\sim 700$ kpc, 
signalled by a reduction in the radial scatter at larger radii. 
The sharp drop of the radial scatter reveals the termination shock, where the outflowing gas motion slows down.

\subsection{Evolution of the Cool Core} 
\label{sec-res-Heat-Cool-Core} 

\begin{figure*} 
\centering 
\includegraphics[width = 1.0 \linewidth]{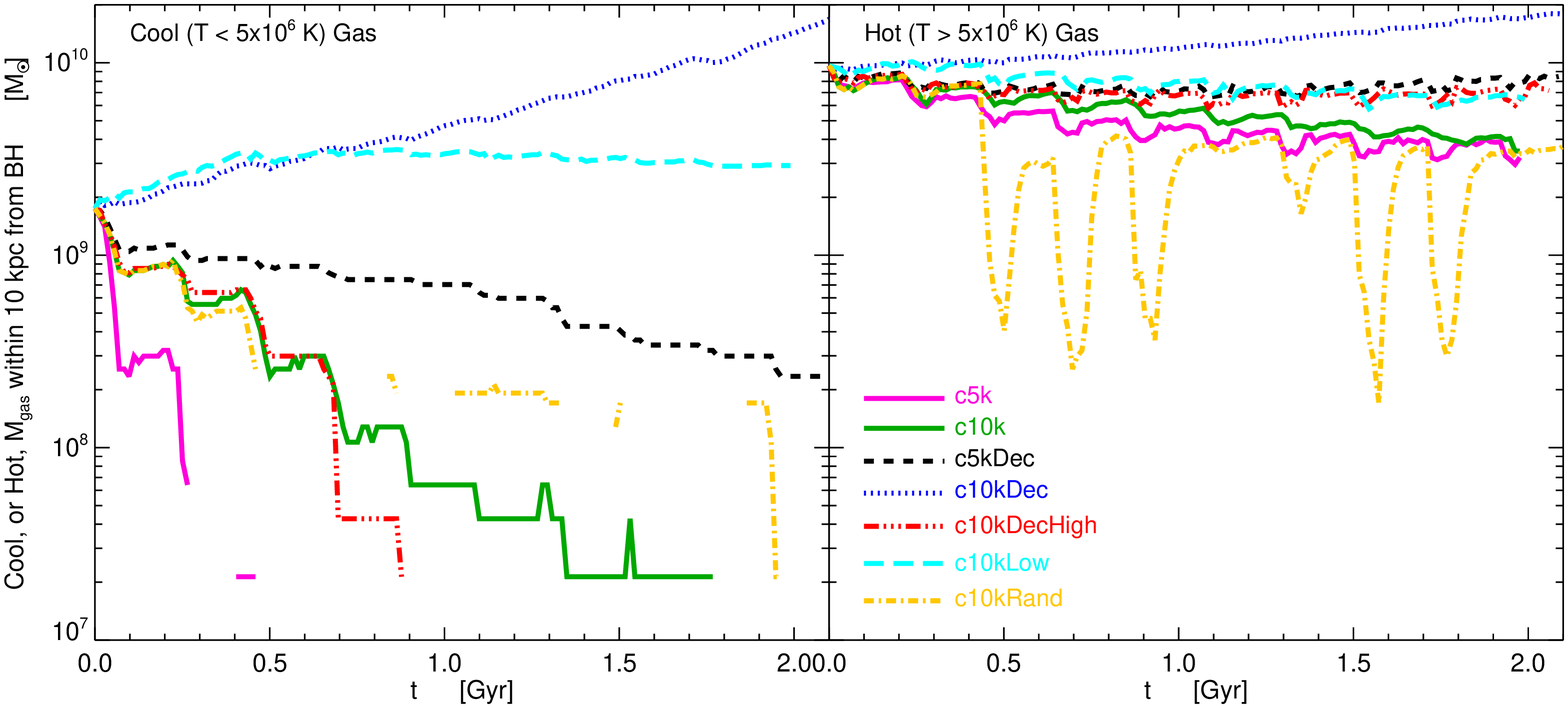} 
\caption{ 
Time evolution of cool and hot gas mass inside distance $r \leq 10 h^{-1}$ kpc 
from the BH at cluster center, in the constant-energy simulations. 
Cool gas of temperature $T \leq 5 \times 10^{6}$ K is shown in the left panel, 
and hot gas of $T > 5 \times 10^{6}$ K at the right. 
The distinguishing colors and linestyles indicate different AGN feedback model runs, 
{\it c5k}: magenta solid, {\it c10k}: green solid, {\it c5kDec}: black dashed, {\it c10kDec}: blue dotted, 
{\it c10kDecHigh}: red dash-dot-dot-dot, {\it c10kLow}: cyan long-dashed, {\it c10kRand}: yellow dash-dot. 
} 
\label{fig-Evol_Gas_Hot_Cold} 
\end{figure*} 

The initial cool core of the cluster is heated up in several simulations with constant-energy output 
from the BH, within a time-period depending on the parameter values of the kinetic feedback model. 
Fig.~\ref{fig-Evol_Gas_Hot_Cold} presents the masses of cool and hot components 
of the gas at the cluster center, lying within distance $r \leq 10 h^{-1}$ kpc from the BH, 
versus evolution time, in the seven constant-energy runs. 
The different curves are for various AGN feedback models, distinguished by the colors and linestyles, 
as labelled and written in the caption. 
Here we define cool gas as that of temperature $T \leq 5 \times 10^{6}$ K which is shown in the left panel, 
and hot gas has $T > 5 \times 10^{6}$ K plotted in the right panel. 
Such a temperature was chosen for the cool/hot distinction based on the IC. 
As seen in Fig.~\ref{fig-IC_Profiles} - blue curve with asterisks, 
the cool core has a median $T = 10^{6}$ K in our initial cluster profiles, 
and the maximum median temperature is $T \sim 2 \times 10^{7}$ K at a larger radii. 

The hot-phase is the dominating component at the center over the cool-phase in all the runs, 
except {\it c10kDec} where the cool-phase dominates after $1.5$ Gyr, reasons for which is elaborated below. 
According to the IC (\S\ref{sec-num-IC}), the cluster inner region consists of 
dense, cold, low-entropy gas (blue curve in Fig.~\ref{fig-IC_Profiles}), 
when the BH is created and allowed to eject feedback energy. 
At $t > 0$ more gas cools, gets denser and sinks to the center, 
tending to enlarge the cool core in the absence of anything else. 

Kinetic feedback from the BH acts as a heating source here. 
A fraction of gas selected from the cool-core reservoir, in particular particles 
within a bi-cone volume inside the smoothing length of the BH, $h_{\rm BH} \sim (1 - 5)$ kpc, 
are chosen and kicked to a higher velocity. 
This drives some of the innermost gas out, 
and ejects it to larger distances in the form of bubble-like outflows (\S\ref{sec-res-outflow}). 
The mass kick out rate in our sub-resolution model, example for run {\it c10k}, is: $\dot{M}_w = 30 M_{\odot}$/yr. 
When the AGN is on for $50$ Myr, 
a total mass of $1.5 \times 10^{9} M_{\odot}$ gas is kicked to a higher speed ($v_w = 10,000$ km/s). 
The total mass outflown from the central $10$ kpc is mostly cool gas. 
The hot (dominating phase), and consequently the total (cool $+$ hot) central gas mass 
remains almost constant with time. 


The kicked fast wind undergoes shock with the slower-moving gas surrounding it, 
which are visible in Fig.~\ref{fig-GasProperties_vs_R_Run14} as the temperature jumps of the shaded areas, 
and were described in \S\ref{sec-res-outflow}. 
The interactions cause the imparted kinetic energy to be thermalized, heating up the cluster cool core. 
It is revealed in the left panel of Fig.~\ref{fig-Evol_Gas_Hot_Cold} 
by the decrease of cool gas mass with time. 
The default output BH power of $\dot{E}_{\rm feed} = 10^{45}$ erg/s is able to heat up the core 
in the cases where the kicked wind particles are always coupled to hydrodynamic interactions: 
{\it c5k} within $0.5$ Gyr, {\it c10k} and {\it c10kRand} by $2$ Gyr. 
For a constant output BH power, $\dot{M}_w$ is proportional to $1 / v_w^2$ 
(Eq.~\ref{eq-Energy-Conservation}). 
So the mass outflow rate is $4$ times higher with $v_w = 5,000$ km/s over $v_w = 10,000$ km/s. 
This causes $4$ times more central gas mass to be kicked in runs {\it c5k} and {\it c5kDec}, 
than in {\it c10k} and {\it c10kDec}. 
This depletes the inner cool gas faster with $v_w = 5,000$ km/s. 

As the extreme opposite case, when wind particles are kicked with $v_w = 10,000$ km/s and 
decoupled from hydrodynamic interactions up to $50$ kpc (run {\it c10kDec}, blue dotted curve), 
it produces increased cooling at the center, when the cool gas mass rises with time. 
This is because the imparted kinetic energy is thermalized at $r > 50$ kpc, 
and does not have any effect at the center. 
The central dense gas continues to cool to $T < 10^{6}$ K here. 
The gas is kicked into wind at the same rate in runs {\it c10k} and {\it c10kDec}, 
however the cool core remains in case {\it c10kDec}. 
This demonstrates that the hydrodynamical interactions of the kicked wind with the surrounding gas 
(occurring in the coupled wind case {\it c10k}) plays an important role in destroying the cool core. 


A higher output BH power $\dot{E}_{\rm feed} = 4 \times 10^{45}$ erg/s 
applied to the same wind-decoupled run ({\it c10kDecHigh}, red dash-dot-dot-dot curve) 
heats up the core completely within $0.9$ Gyr. 
The wind-decoupled case with default output BH power and 
lower velocity $v_w = 5,000$ km/s (run {\it c5kDec}, black dashed curve) 
depletes the cool central gas ($10$ times reduction in cool mass in $2$ Gyr), 
albeit rather slowly, since $\dot{M}_w$ is $4$ times higher than in run {\it c10kDec}. 

\begin{figure*} 
\centering 
\includegraphics[width = 0.7 \linewidth]{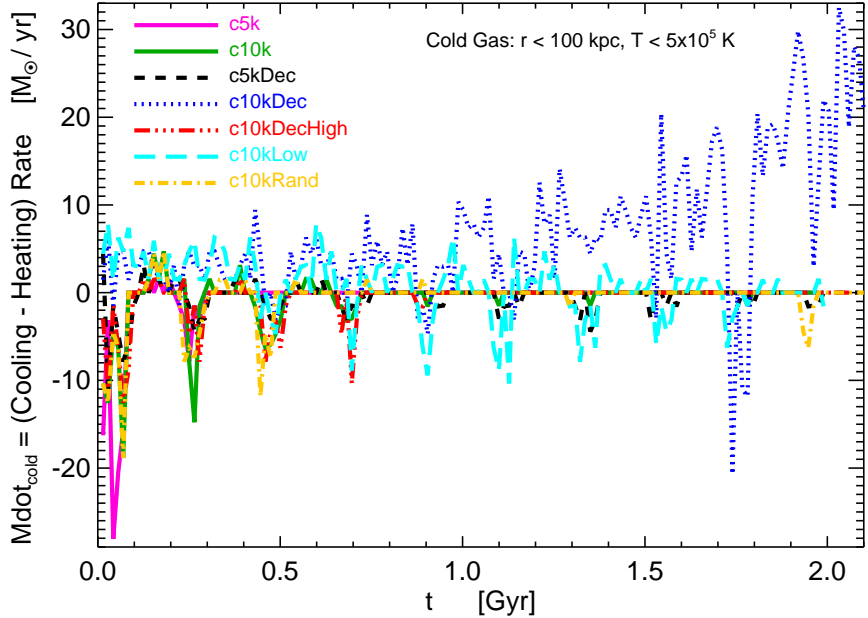} 
\caption{ 
Time evolution of mass (cooling - heating) rate in the constant-energy simulations, 
considering cold gas at $r \leq 100$ kpc and of temperature $T \leq 5 \times 10^{5}$ K. 
Positive values indicate net cooling of the gas, and negative values indicate net heating. 
The distinguishing colors and linestyles indicate different AGN feedback models, 
{\it c5k}: magenta solid, {\it c10k}: green solid, {\it c5kDec}: black dashed, {\it c10kDec}: blue dotted, 
{\it c10kDecHigh}: red dash-dot-dot-dot, {\it c10kLow}: cyan long-dashed, {\it c10kRand}: yellow dash-dot. 
} 
\label{fig-Evol_Mdot_Cold} 
\end{figure*} 

Fig.~\ref{fig-Evol_Mdot_Cold} shows the time evolution of mass (cooling - heating) rate of the gas 
in the constant-energy runs. 
In computing this net cooling rate, 
we consider gas lying inside distance $r < 100$ kpc and of temperature $T < 5 \times 10^{5}$ K as cold. 
Positive values indicate net cooling of the gas, and negative values indicate net heating. 
The periodic negative spikes indicate the AGN outburst heating at the rate $(10 - 30) M_{\odot}$/yr. 
During AGN-off periods there is cooling at the rate $< \sim 10 M_{\odot}$/yr, 
except run {\it c10kDec} where the cooling rate is higher. 

The lower output BH power $\dot{E}_{\rm feed} = 2 \times 10^{44}$ erg/s having always coupled wind 
(run {\it c10kLow}, cyan long-dashed curve) maintains the cool core, 
with the inner cool gas mass (Fig.~\ref{fig-Evol_Gas_Hot_Cold}) remaining almost constant over time. 
In this case, the smaller amount of energy injected allows cooling to occur 
at a rate few $M_{\odot}$/yr even after $1$ Gyr, which keeps the cluster core in nearly thermal equilibrium. 

We find that if the direction of imparting kinetic feedback changes randomly 
from one activity cycle of BH to another (run {\it c10kRand}, yellow dash-dot curve), 
it has a significant impact on the central hot gas mass. 
It is revealed by the decrease of inner hot gas mass by $10$ times or more 
(large spikes in the right panel of Fig.~\ref{fig-Evol_Gas_Hot_Cold}) during certain activity cycles, 
while returning back to the original value after $100$ Myr. 
Here the hot gas is expelled outside $10$ kpc from the central BH, 
by strong kinetic feedback, which is much more efficient 
with the change of the feedback direction between duty cycles of BH. 

\begin{figure*} 
\centering 
\includegraphics[width = 1.0 \linewidth]{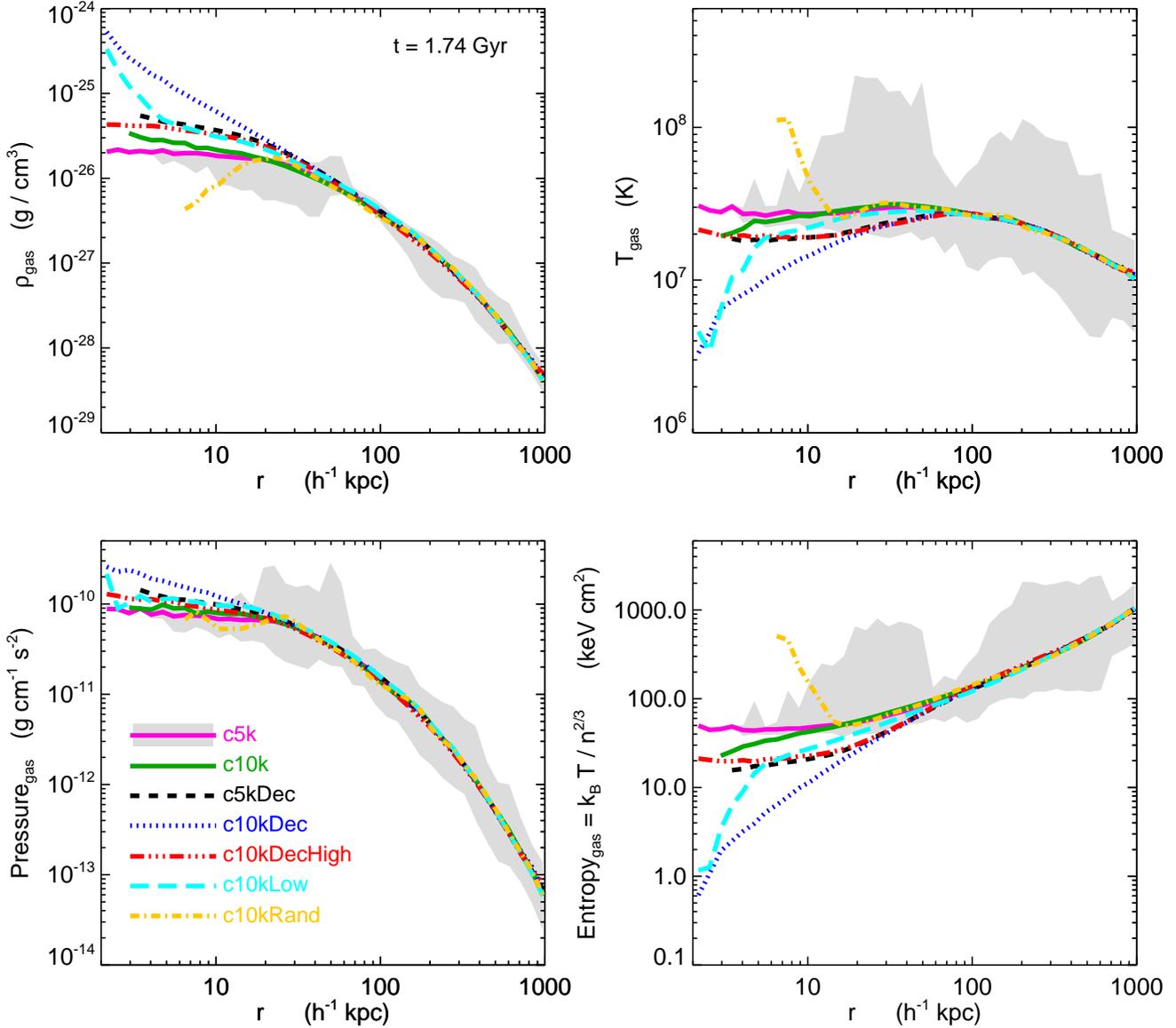} 
\caption{ 
Radial profiles of gas properties at an evolution time of $1.74$ Gyr in the constant-energy simulations: 
density in the top-left panel, temperature at the top-right, 
pressure at the bottom-left, and entropy in the bottom-right panel. 
The plotted curves denote the median quantity in radial bins of each run. 
The distinguishing colors and linestyles indicate runs with different AGN feedback models, 
{\it c5k}: magenta solid, {\it c10k}: green solid, {\it c5kDec}: black dashed, {\it c10kDec}: blue dotted, 
{\it c10kDecHigh}: red dash-dot-dot-dot, {\it c10kLow}: cyan long-dashed, {\it c10kRand}: yellow dash-dot. 
The grey shaded areas enclose the $90$th percentiles above and below the median of run {\it c5k}, 
as a representative of the radial scatter. 
} 
\label{fig-GasProperties_vs_R_9runs} 
\end{figure*} 

Fig.~\ref{fig-GasProperties_vs_R_9runs} displays the radial profiles of gas properties 
in the constant-energy runs at an evolution time $t = 1.74$ Gyr. 
The radius is computed by the distance from the BH location, and we have checked that the BH 
remains within $0.6$ kpc from the cluster center, which is the origin of our spherical coordinates. 
The profiles of the seven runs are indistinguishable from each other at larger radii: 
$r > 20$ kpc for density (top-left panel) and pressure (bottom-left); 
$r > 70$ kpc for temperature (top-right) and entropy (bottom-right). 
Kinetic feedback from the BH affects the gas properties at only the regions inner to these radii. 
It expels some central gas in the form of high-velocity outflows, 
which thermalize their energies and shock-heat the core. 
The core heating is exhibited by the median temperature remaining high $T > 10^{7}$ K in the inner regions. 
This corresponds to a central entropy flattening, $S = k_{\rm B} T / n^{2/3} > 15$ keV cm$^2$. 

The cluster core is heated to $T > 10^{7}$ K and having a flat entropy profile, 
by a time $t \sim 1.8$ Gyr in five runs: 
{\it c5k}, {\it c10k}, {\it c5kDec}, {\it c10kDecHigh}, and {\it c10kRand}. 
Kinetic feedback from the BH is less efficient in two runs, 
where the initial cool core ($T \leq 5 \times 10^{6}$ K) remains: 
{\it c10kDec} (blue dotted curve in Fig.~\ref{fig-GasProperties_vs_R_9runs}) 
where the kinetic energy is thermalized too far away from the center, 
and {\it c10kLow} (cyan long-dashed) where the low output power keeps the cluster core in nearly thermal equilibrium. 
These cool-core clusters have a $10 -100$ times higher central density than the cases where the core is heated. 
The pressure variation between different models is smaller than the variations in other gas properties, 
because pressure mostly depends on the background potential which is static in our simulations.

\subsection{Resolution Study} 
\label{sec-res-Resolution} 

\begin{figure*} 
\centering 
\includegraphics[width = 1.0 \linewidth]{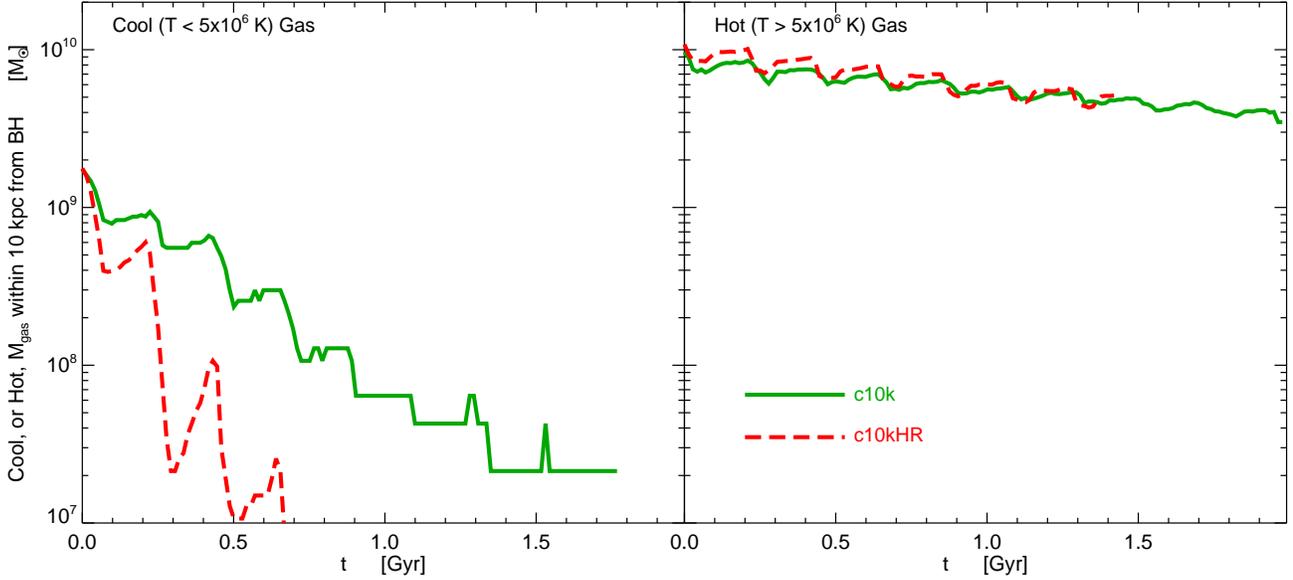} 
\caption{ 
Time evolution of cool ($T \leq 5 \times 10^{6}$ K, left panel) and hot ($T > 5 \times 10^{6}$ K, right panel) 
gas mass inside $10 h^{-1}$ kpc from the BH at cluster center, in the varying resolution simulations; 
the default {\it c10k}: green solid curve, and the high-resolution run {\it c10kHR}: red dashed curve. 
} 
\label{fig-Evol_Gas_Hot_Cold_HiRes} 
\end{figure*} 

We perform a higher resolution simulation {\it c10kHR} 
with $10$ times more number of particles (Table~\ref{Table-Galaxies}) than our default resolution cases. 
We chose the constant-energy setup for our resolution study because it is the simplest with no star-formation 
(i.e. free from resolution dependence of the star-formation model), 
thus here the resolution effect of the AGN energy injection scheme can be better identified. 
{\it c10kHR} employs the same model parameters as run {\it c10k}. 
Fig.~\ref{fig-Evol_Gas_Hot_Cold_HiRes} shows the masses of cool and hot components 
of the gas at the cluster center, lying within distance $r \leq 10 h^{-1}$ kpc from the BH, 
versus evolution time, in these two different resolution runs. 
We find that the cluster cool core is heated up faster in the high-resolution case, 
as seen in the left panel of Fig.~\ref{fig-Evol_Gas_Hot_Cold_HiRes}. 
The inner cool gas mass decreases with time and is depleted within $0.7$ Gyr in run {\it c10kHR} (red dashed curve), 
which is almost half the time needed in the lower resolution case {\it c10k} (green solid curve). 

\begin{figure*} 
\centering 
\includegraphics[width = 1.02 \linewidth]{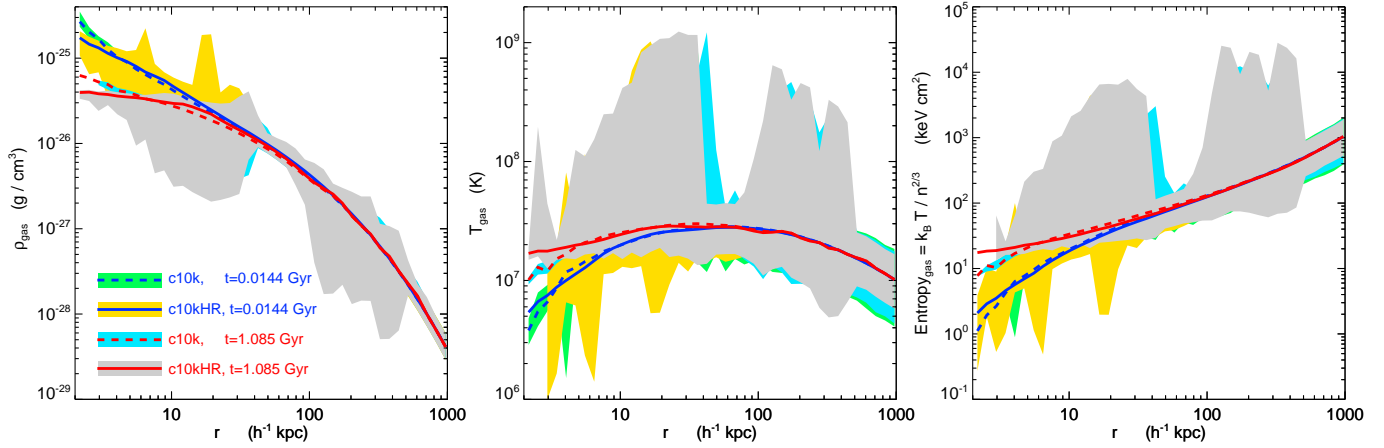} 
\caption{ 
Radial profiles of gas properties at two different epochs in the varying resolution simulations; 
the default {\it c10k}: dashed curves, and the high-resolution {\it c10kHR}: solid curves. 
Left panel is gas density, middle panel is temperature, and right panel is entropy. 
The plotted times are: 
$t = 0.01$ Gyr (blue curves with green and yellow shaded areas) showing the first outflow propagation, 
$t = 1.08$ Gyr (red curves with cyan and grey shaded areas) a later time AGN-on active epoch. 
The curves denote the median quantity in radial bins, and 
the shaded areas enclose the $90$th percentiles above and below the median, showing the radial scatter. 
Note the different color scheme in this figure compared to other figures. 
} 
\label{fig-GasProperties_vs_R_HiRes} 
\end{figure*} 

Fig.~\ref{fig-GasProperties_vs_R_HiRes} presents the radial profiles of gas properties 
at two different epochs in runs {\it c10k} and {\it c10kHR}. 
The varying resolution brings a difference in the median profiles at $r < 10$ kpc. 
Bubble-like outflows are visible as rise in the radial scatter (shaded areas): 
hot (middle panel) reaching temperature as high as $10^{9}$ K, high-entropy (right) and low-density (left). 
The pressure profile remains the same in the different resolution runs, and we do not show it here. 
The blue curves ($t = 0.01$ Gyr) depict an epoch of the first active period of the BH. 
The red curves ($t = 1.08$ Gyr) are a later time AGN-on active epoch. 
Comparing the cyan and grey shaded areas, we see that the inner bubble 
starts from a smaller radii at high-resolution, and extends up to a shorter distance. 
The outer bubble signatures, extending over $100 - 600$ kpc, remain the same when resolution is varied. 

Comparing between the solid and dashed curves at each epoch in Fig.~\ref{fig-GasProperties_vs_R_HiRes}, 
we see that the cluster core is heated more at high-resolution, entropy is higher, and the density is lower. 
However the differences are not large and remain within a factor of $2$. 
Moreover, the profiles at the two resolutions show an excellent degree of convergence 
at length scales larger than $\approx 10$ kpc. 

We thus find that convergence is not reached strictly in our simulations. 
Our implementation of kinetic AGN feedback is more effective at higher resolution. 
Full convergence cannot be expected using the same parameter values 
within the framework of our sub-resolution models. 
This is because feedback energy is distributed to the local gas, 
and higher gas densities are resolved in the high-resolution case. 
As discussed in \citet{Barai14} (\S4, paragraph 9), the lack of convergence implies that different 
model parameters (e.g. $\epsilon_f, v_w$) are required at different resolution 
in cosmological simulations to have the same fit to 
observational correlations (e.g. BH mass versus host galaxy stellar velocity dispersion). 

In a contemporary work investigating numerical resolution, \citet{Bourne15} studied the behaviour of 
AGN feedback when force and mass resolution are varied. 
They found that an {\it increase} of resolution results in a {\it less} effective ejection of the gas by AGN feedback. 
The result we obtain here is at variance with theirs. 
We however note that both the resolution and coupling of the AGN energy to the gas 
is quite different in our work with respect to \citet{Bourne15}. 
They simulated the interaction of an AGN outflow with the clumpy turbulent gas in the inner part of the host galaxy 
on length scales from $0.1$ to $1$ kpc, which is smaller than ours ($> 2$ kpc). 
Feedback-resistant high-density clumps are washed out at low effective resolutions in their simulations, 
whereas such clumps never form in our simulations. 
Our mass resolution is typical of that of a cosmological simulation of galaxy clusters, 
thus our gas particle mass is at least an order of magnitude higher than theirs. 
Moreover, in that paper the coupling of energy is done by injecting thermal energy. 
Hence in our opinion, the point of numerical convergence is still open, 
since it can easily depend on both the simulation force and length scales 
and on the precise scheme adopted for the energy deposition onto the gas.

\section{Results: BH Growth and Energy Output from BH Accretion} 
\label{sec-results-BH-Accr-Feedback} 

A BH is seeded at the cluster center at $t = 0$ according to our IC (\S\ref{sec-num-IC}), 
of initial mass $M_{\rm BH} = 10^5 M_{\odot}$ (\S\ref{sec-num-Implement}). 
In the next eight simulations with cooling, SF and BH accretion, the BH grows in mass 
by accreting gas from its surroundings, and consequently outputs feedback energy (\S\ref{sec-num-BH-Accr-Feed}). 
We find that kinetic feedback-induced gas outflows are created, 
at the times when the BH undergoes a high accretion rate. 
The cluster core is heated eventually when using higher values of the feedback efficiency parameter.

\subsection{Black Hole Accretion and Mass Growth} 
\label{sec-res-BH-Growth} 

\begin{figure*} 
\centering 
\includegraphics[width = 0.8 \linewidth]{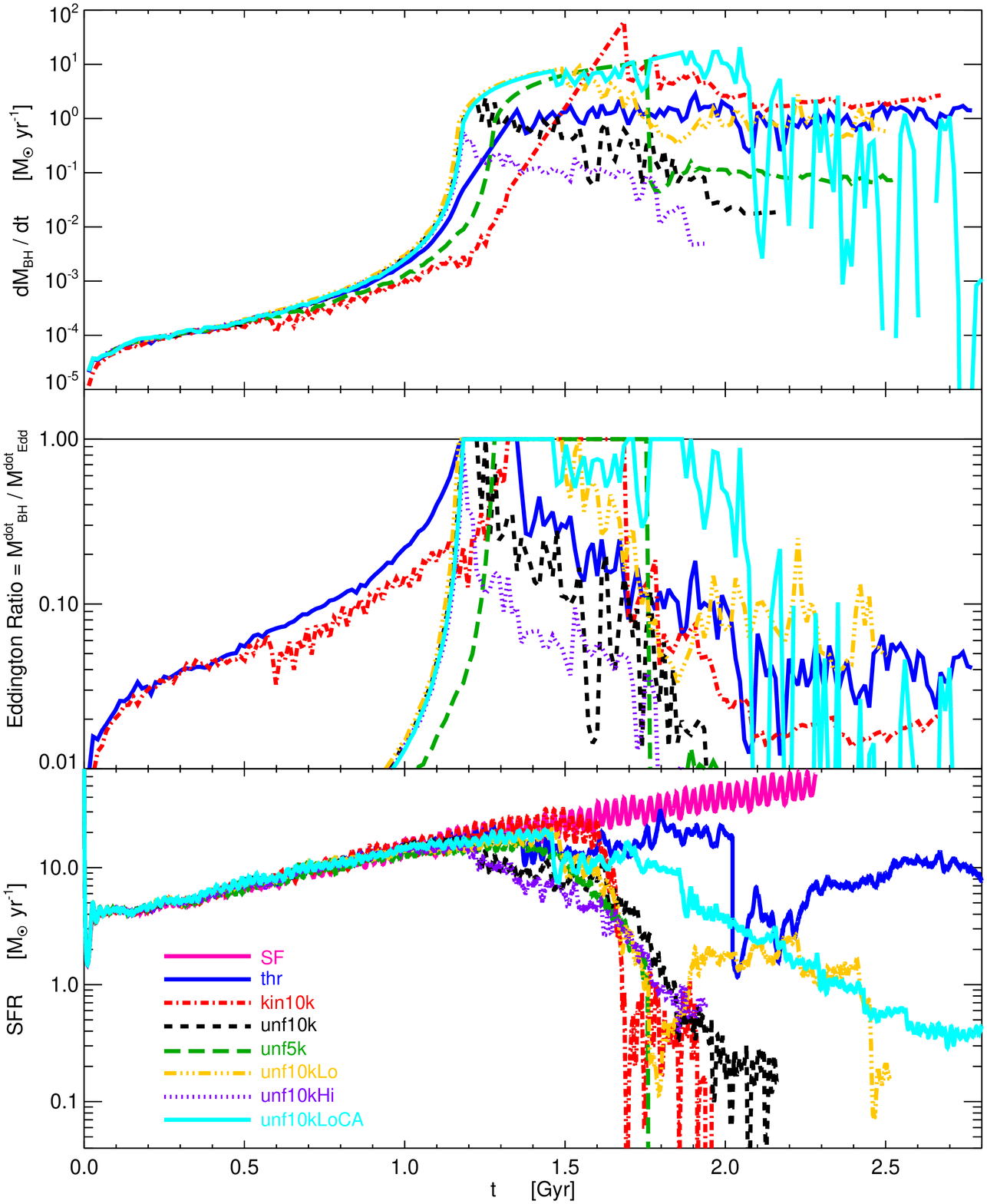} 
\caption{ 
Evolution with time of BH mass accretion rate (top panel), Eddington ratio (middle panel), 
and total star formation rate (bottom panel), in the simulations with cooling, SF and BH growth. 
The different colors and linestyles discriminate AGN feedback models as labelled in the bottom panel, 
{\it SF}: magenta solid, {\it thr}: blue solid, {\it kin10k}: red dash-dot, {\it unf10k}: black dashed, 
{\it unf5k}: green long-dashed, {\it unf10kLo}: yellow dash-dot-dot-dot, {\it unf10kHi}: purple dotted, 
{\it unf10kLoCA}: cyan solid. 
} 
\label{fig-BHAccr_StarForm_Rate} 
\end{figure*} 

The time evolution of BH mass accretion rate ($\dot{M}_{\rm BH}$) 
is presented in Fig.~\ref{fig-BHAccr_StarForm_Rate}, top panel. 
The middle panel displays the Eddington ratio $= \dot{M}_{\rm BH} / \dot{M}_{\rm Edd}$. 
The eight runs with different AGN feedback models are distinguished 
by the colors and linestyles, as labelled in the bottom panel and caption. 
At $t = 0$, the $\dot{M}_{\rm BH} = 10^{-5} M_{\odot}$/yr is very low and rises slowly with time. 
From $t \sim 1.1$ Gyr, the $\dot{M}_{\rm BH}$ has a power-law increase, 
corresponding to the Eddington-limited growth of the BH (where Eddington ratio $= 1$), 
up to a time $\sim 1.5$ Gyr, to reach a peak of $\dot{M}_{\rm BH}$. 
As an example, the peak $\dot{M}_{\rm BH}$ of $5 M_{\odot}$/yr in run {\it unf10k} occurs at $1.25$ Gyr. 

Kinetic feedback (run {\it kin10k}, red dash-dot curve) produces a $50$ times larger 
peak $\dot{M}_{\rm BH}$ than thermal feedback ({\it thr}, blue solid) using the same $\epsilon_f = 0.02$. 
After its peak, the $\dot{M}_{\rm BH}$ reduces in the runs with kinetic feedback 
({\it kin10k} and the {\it unf} cases), because significant amounts of gas are ejected 
from the center and/or heated up, limiting the gas feeding onto the BH. 
While in the thermal feedback run {\it thr}, the $\dot{M}_{\rm BH}$ is almost constant after the peak, 
because cold gas remains at the center here, retaining the cool core. 

There are fluctuations in the $\dot{M}_{\rm BH}$, 
whereby it increases or decreases by a factor of up to $10$ in $0.02$ Gyr. 
A noteworthy feature is the huge variability of $\dot{M}_{\rm BH}$ in run {\it unf10kLoCA} at $t > 2$ Gyr. 
Here the inclusion of cold gas accretion results in $\dot{M}_{\rm BH}$ fluctuating 
by a factor of up to $10^4$, with a periodicity of $100$ Myr. 


\begin{figure*} 
\centering 
\includegraphics[width = 1.0 \linewidth]{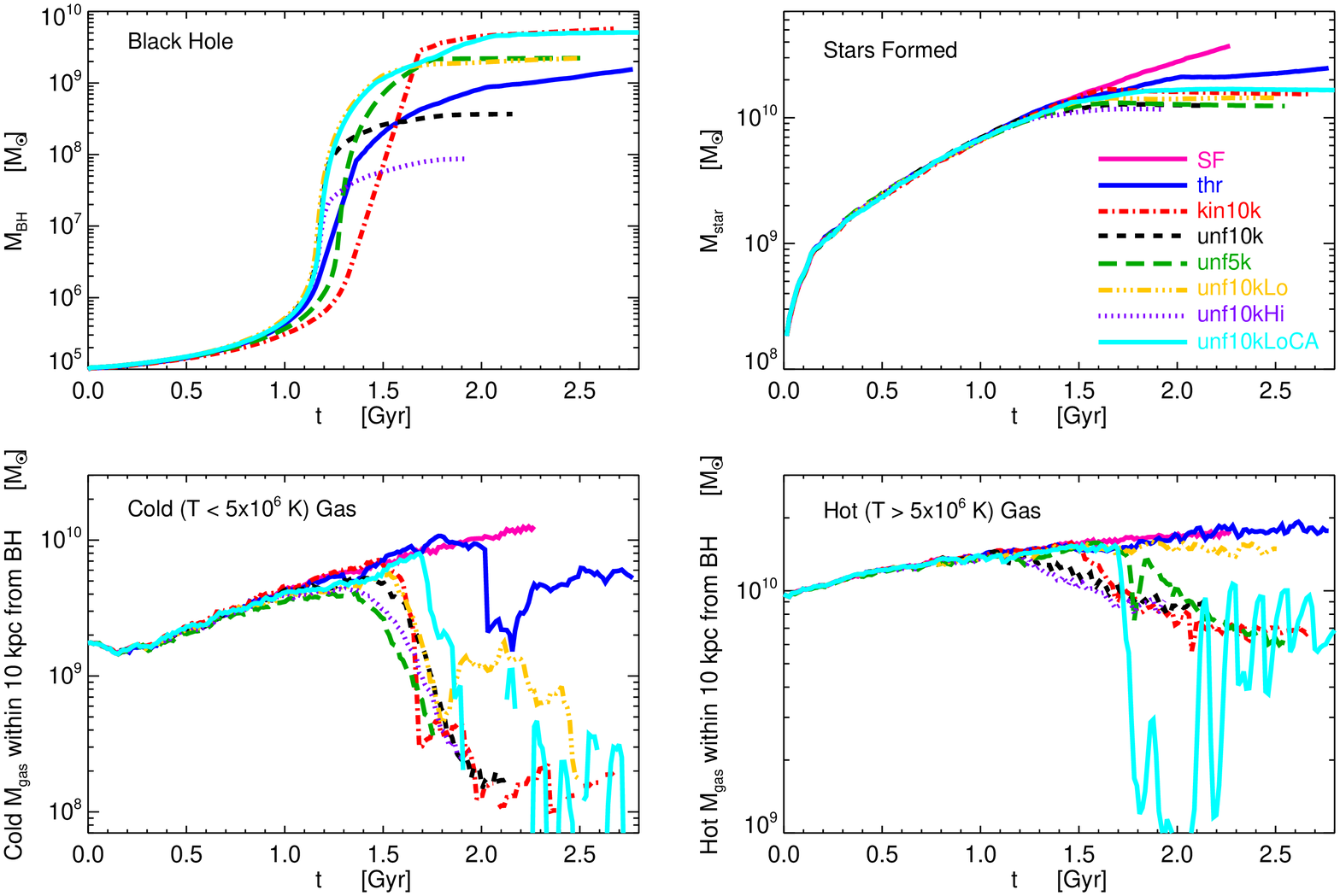} 
\caption{ 
Time evolution of BH, stellar, and central gas masses in the simulations with cooling, SF and BH growth: 
BH mass (top-left panel), 
mass of stars formed from gas via SF (top-right), 
cool gas mass of temperature $T \leq 5 \times 10^{6}$ K inside distance $r \leq 10 h^{-1}$ kpc 
from the BH location (bottom-left), 
and hot gas mass of $T > 5 \times 10^{6}$ K inside $10 h^{-1}$ kpc from BH (bottom-right). 
The different colors and linestyles discriminate AGN feedback models as labelled in the top-right panel: 
{\it SF}: magenta solid, {\it thr}: blue solid, {\it kin10k}: red dash-dot, {\it unf10k}: black dashed, 
{\it unf5k}: green long-dashed, {\it unf10kLo}: yellow dash-dot-dot-dot, {\it unf10kHi}: purple dotted, 
{\it unf10kLoCA}: cyan solid. 
} 
\label{fig-Evol_Gas_Star_BH} 
\end{figure*} 

The BH mass versus evolution time is plotted in Fig.~\ref{fig-Evol_Gas_Star_BH}, top-left panel. 
All the feedback models have the BH growing in a qualitatively similar manner. 
Starting from the given seed mass of $10^5 M_{\odot}$, each BH has an exponential growth. 
Over the time range $1$ Gyr to $1.5$ Gyr, its mass increases by a factor $10^3$ to $10^5$. 
After $t \sim 1.5$ Gyr, $M_{\rm BH}$ comes to an almost steady state, having a very slow subsequent growth. 
The final $M_{\rm BH}$ reached at $2$ Gyr depends on the model: 
e.g.~$3 \times 10^8 M_{\odot}$ in run {\it unf10k}. 
$M_{\rm BH}$ is inversely proportional to $\epsilon_f$ in the unified feedback model 
(runs {\it unf10kLo}: yellow dash-dot-dot-dot curve, {\it unf10k}: black dashed, {\it unf10kHi}: purple dotted; 
using $\epsilon_f = 0.002, 0.02, 0.2$). 
This is due to self regulation of the BH growth \citep[see also e.g.][]{Booth09}, and is analytically expected 
when the BH feedback energy is derived from $\dot{M}_{\rm BH}$ (Eq.~\ref{eq-Edot-Feed}). 
A higher $\epsilon_f$ imparts a stronger feedback ejecting out more central gas, 
and yields a less-massive BH than a lower $\epsilon_f$. 

The BH mass is also inversely proportional to the kick velocity $v_w$ for unified feedback 
(runs {\it unf5k}: green long-dashed, and {\it unf10k}: black dashed; using $v_w = 5,000$ and $10,000$ km/s), 
which is contrary to naive expectations. 
Note that \citet{Barai14} saw a direct proportionality of $M_{\rm BH}$ with $v_w$, 
for kinetic feedback in isolated galaxies and mergers. 
This is analytically expected because on increasing $v_w$, 
$\dot{M}_w$ decreases (inverse proportionality in Eq.~\ref{eq-MdotW-EDW}). 
The reduced mass outflow rate of kinetic feedback ejects out less gas. 
Hence, more gas is available for accreting onto the BH, which grows more massive. 
We posit that the discrepancy between the two studies arises from the simulation setup. 
Here the BH kicks out central gas within a cluster atmosphere, and at later times some of the ejected gas falls back in, 
after slowing down by interacting with the further-out intracluster gas. 
Whereas in \citet{Barai14}, once the kicked gas escaped the galaxy, it never came back. 



\subsection{Formation of Stars} 
\label{sec-res-SF} 

Stars form in the isolated cluster from dense gas according to the SF prescription 
described in \S\ref{sec-numerical}. 
The SF occurs within a few kpc region at the cluster center where the gas inflows and cools, 
forming an initial cool core. 
The presence of a BH decreases the central gas content, as some gas is accreted in, 
some gas is ejected by feedback, and some gas is heated; which consequently quenches SF. 

The total star formation rate versus time is displayed in the bottom panel of 
Fig.~\ref{fig-BHAccr_StarForm_Rate}, for eight runs with different AGN feedback models. 
At $t = 0$ there is an initial burst of SF at a rate $80 M_{\odot}$/yr, because of our simulation IC: 
the cluster already has substantial amounts of cold, dense gas in the form of a cool core, 
which readily form stars when SF is allowed in the model. 
It reduces by $40$ times soon afterward, because of reduced central gas, 
which has depleted by the initial SF burst. 
The SFR increases almost linearly with time from $0.2$ Gyr, 
as more gas cools, gets dense and is converted to stars. 
There are periodic fluctuations in the SFR, when it would decrease by a factor of $\sim 2$, 
occurring because of supernovae feedback in the stellar evolution model. 
All the runs show this rising SFR initially, 
which continues to increase with time in the {\it SF} case without a BH. 

The models start to deviate from $1.2$ Gyr onwards, 
when the BH has grown to a high-mass and generates enough feedback suppressing SF. 
Thermal feedback (run {\it thr}, blue solid curve) quenches SF at $t > 1.3$ Gyr, 
the SFR is lower than the {\it SF} case (magenta solid curve) by up to a factor of $10$. 
Kinetic feedback ({\it kin10k} and the {\it unf} cases) 
causes a greater suppression of SFR up to few $100$ times lower than the {\it SF} run, 
because significant amounts of gas are ejected from the center. 
As an example, in run {\it unf10k} (black dashed curve) the SFR is reduced 
from $15 M_{\odot}$/yr at $1.1$ Gyr to $0.2 M_{\odot}$/yr at $2$ Gyr. 

Fig.~\ref{fig-Evol_Gas_Star_BH}, top-right panel displays the time evolution of 
mass of stars formed from gas via SF. 
Gas is continuously converted to stars by the rising SFR initially, for all the AGN models. 
The stellar mass hence increases with simulation time up to $1.2$ Gyr, 
by when a total $M_{\rm star} = 10^{10} M_{\odot}$ of stars has been formed, 
an amount which is comparable to the hot gas mass in the central $10$ kpc of the cluster. 
After $1.2$ Gyr, $M_{\rm star}$ continues to rise with time in the {\it SF} case. 
$M_{\rm star}$ also increases with thermal feedback albeit with a smaller slope. 
While in the runs with kinetic feedback $M_{\rm star}$ remains constant 
as no more new stars are formed by the hugely quenched SFR.

\subsection{Evolution of the Cool Core by Growing BH Feedback} 
\label{sec-res-Heat-Cool-Core-BH} 

The initial cool core of the cluster is heated, by energy feedback from the growing BH, 
to different extents depending on the model parameter values. 
The bottom row of Fig.~\ref{fig-Evol_Gas_Star_BH} presents the masses of cool and hot components 
of gas within a distance $r \leq 10 h^{-1}$ kpc from the BH location, versus evolution time. 
Here cool gas is that of temperature $T \leq 5 \times 10^{6}$ K shown in the bottom-left, 
and hot gas has $T > 5 \times 10^{6}$ K plotted in the bottom-right panel. 

The hot phase dominates at the cluster center over the cool phase in all the runs, 
except {\it SF} (magenta solid curve) where the two phases become comparable after $2$ Gyr. 
The hot gas mass increases slowly with time, initially for all the models 
to reach $M_{\rm hot} \sim 1.3 \times 10^{10} M_{\odot}$ at $1.2$ Gyr. 
This rising $M_{\rm hot}$ is caused by the inflow of hotter gas from the surroundings, 
because of gas depletion near the center by cooling-induced SF. 
$M_{\rm hot}$ continues to rise for three runs: {\it SF}, {\it thr}, {\it unf10kLo}, 
which has either no or low BH feedback. 
While $M_{\rm hot}$ reduces subsequently for five runs: 
{\it kin10k}, {\it unf10k}, {\it unf5k}, {\it unf10kHi}, {\it unf10kLoCA},  
decreasing by $2$ times from $\sim 1.3$ Gyr to $2.5$ Gyr, 
showing the influence of strong kinetic feedback in these cases where significant amount of central gas outflows. 

A fraction of gas cools with time, gets denser, and sinks to the cluster inner region, forming a cool core. 
Initially the BH is small and does not exert enough feedback to influence the gas. 
Therefore the cool gas mass within the inner $10 h^{-1}$ kpc 
increases with time up to $M_{\rm cool} \sim 5 \times 10^{9} M_{\odot}$ at $1.4$ Gyr. 
By this time the BH has grown to $M_{\rm BH} \geq 5 \times 10^7 M_{\odot}$. 
Subsequently the BH exerts substantial kinetic feedback energy heating the cool core with some parameters, 
revealed in the bottom-left panel of Fig.~\ref{fig-Evol_Gas_Star_BH} by the decrease of cool gas mass with time. 

All the models with kinetic feedback ({\it kin10k} and {\it unf} runs) heat the core, 
reducing the cool gas to $M_{\rm cool} \sim 2 \times 10^{8} M_{\odot}$ by $1.8$ Gyr. 
In the case {\it unf10kLo} (yellow dash-dot-dot-dot curve), $M_{\rm cool}$ increases again 
forming a local peak between the time range $1.8 - 2.5$ Gyr, because it uses the 
smallest $\epsilon_f = 0.002$ and the low feedback allows some gas to cool at late times. 
With no BH feedback, run {\it SF} produces increased cooling at the center, 
when the cool gas mass monotonically rises with time. 
Thermal feedback ({\it thr}, blue solid curve) is able to deplete some of the cool central gas, 
which reduces by $5$ times at $2$ Gyr, but increases again later, showing that fraction of the cool core remains. 

Run {\it unf10kLoCA} presents features similar to the variability of $\dot{M}_{\rm BH}$ 
(\S\ref{sec-res-BH-Growth}) at $t > 2$ Gyr. 
Cold gas accretion results in the cool and hot gas mass fluctuating 
by a factor of up to $10$, with a periodicity of $100$ Myr.

\begin{figure*} 
\centering 
$ 
\begin{array}{c} 
\includegraphics[width = 0.9 \linewidth]{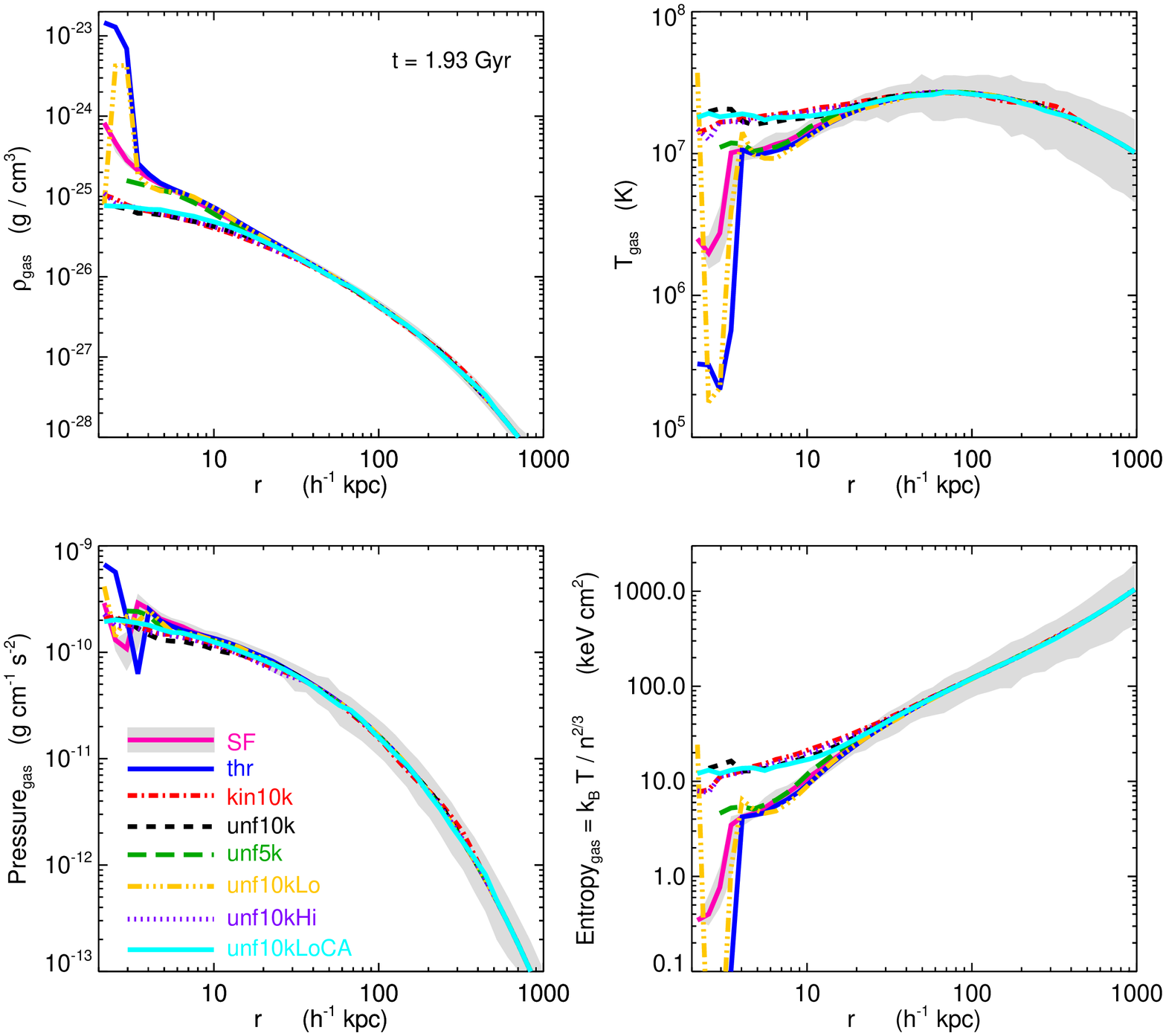} \\ 
\includegraphics[width = 0.9 \linewidth]{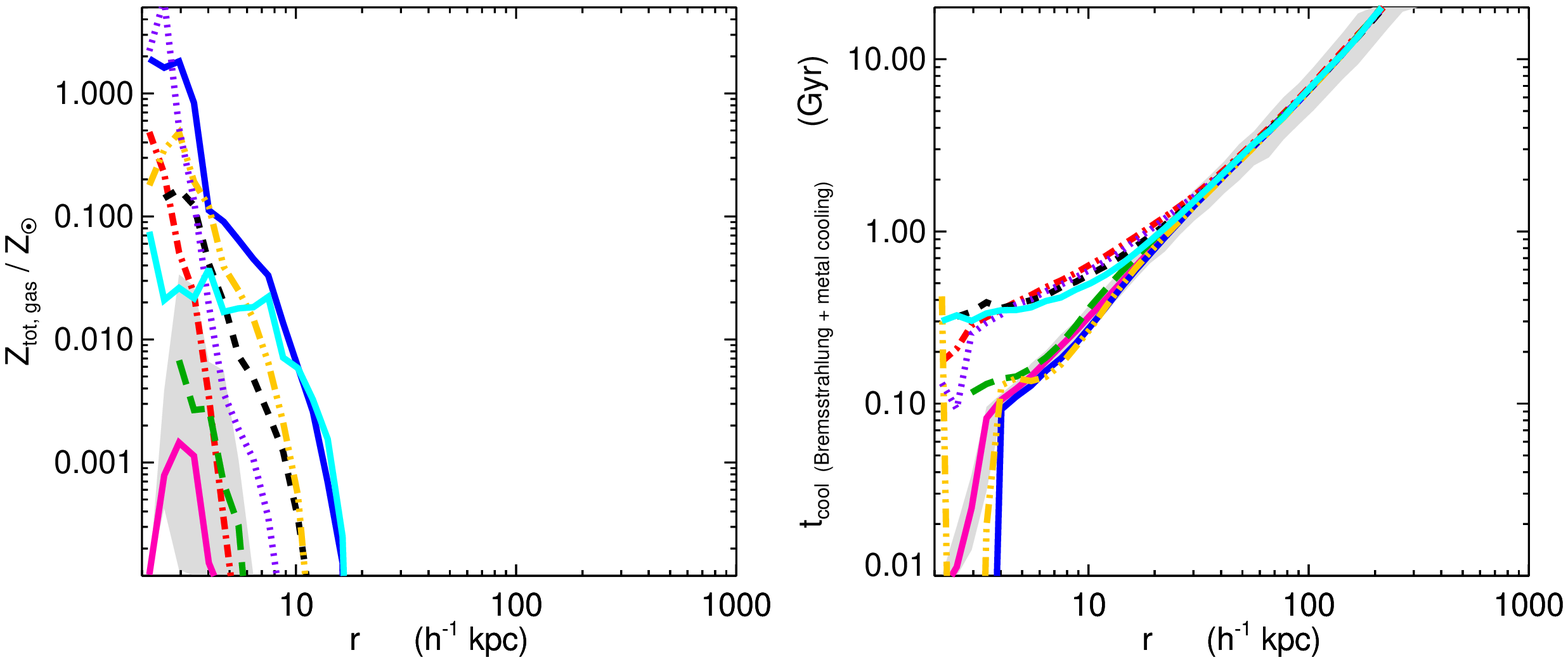} 
\end{array} 
$ 
\caption{ 
Radial profiles of gas properties in the simulations with cooling, SF and BH growth, 
at an evolution time of $1.93$ Gyr: 
density (top-left panel), temperature (top-right), pressure (middle-left), entropy (middle-right), 
total metallicity (bottom-left), and cooling time for Bremsstrahlung and metal line cooling (bottom-right panel). 
The plotted curves denote the median quantity in radial bins of each run. 
The distinguishing colors and linestyles indicate runs with different AGN feedback models, 
as labelled in the middle-left panel, 
{\it SF}: magenta solid, {\it thr}: blue solid, {\it kin10k}: red dash-dot, {\it unf10k}: black dashed, 
{\it unf5k}: green long-dashed, {\it unf10kLo}: yellow dash-dot-dot-dot, {\it unf10kHi}: purple dotted, 
{\it unf10kLoCA}: cyan solid. 
The grey shaded areas enclose the $90$th percentiles above and below the median of run {\it SF}, 
as a representative of the radial scatter. 
} 
\label{fig-GasProperties_vs_R_7runs} 
\end{figure*} 

Fig.~\ref{fig-GasProperties_vs_R_7runs} displays the radial profiles of gas properties 
in the simulations with cooling, SF and BH growth, at an evolution time $t = 1.93$ Gyr. 
The profiles of the eight runs are indistinguishable from each other at larger radii $r > 20$ kpc. 
BH accretion and feedback influence the gas properties in the inner regions ($r < 20$ kpc) only. 
Some central gas is expelled as high-velocity outflows by kinetic BH feedback, 
which thermalize their energies and shock heat the core. 
The core heating is exhibited by the median temperature (top-right panel) 
remaining high $T > 10^{7}$ K in the inner regions. 
This corresponds to a central entropy flattening, $S = k_{\rm B} T / n^{2/3} > 5$ keV cm$^2$ (middle-right panel). 

The cluster core is heated to $T > 10^{7}$ K with a flat entropy profile, by a time $t = 1.9$ Gyr in five runs: 
{\it kin10k}, {\it unf10k}, {\it unf5k}, {\it unf10kHi}, {\it unf10kLoCA}. 
The initial cool core (central $T \leq 10^{6}$ K) remains in three runs: 
{\it SF} where there is no BH, {\it thr} since thermal feedback is less effective, and 
{\it unf10kLo} where the kinetic feedback power is low because of the smallest $\epsilon_f = 0.002$ value used. 
These cool-core clusters have a $10 -100$ times higher central density, 
and up to $10$ times higher central pressure, than the cases where the core is heated up. 

We compute the total gas metallicity, $Z_{\rm gas}$, 
as the ratio of all metal mass to the total gas mass for each gas particle. 
Abundance ratios are expressed in terms of the Solar metallicity, 
which is $Z_{\odot} = 0.0122$ (mass fraction of all metals in Sun) derived from the compilation by \citet{Asplund05}. 
The gas metallicity radial profiles are plotted in the bottom-left panel of Fig.~\ref{fig-GasProperties_vs_R_7runs}. 
The metallicity profiles are very steep, since metals are produced in the star-forming gas, 
concentrated inside the central $10$ kpc. 
The median metal abundance at $2$ kpc reaches a few times Solar in runs {\it thr} and {\it unf10kHi}, 
and reduces by 4 orders of magnitude at $10$ kpc. 
We checked that the run {\it SF} (where there is no BH, and hence the largest star formation rate) generates 
the highest median metallicity among all the runs of $10 Z_{\odot}$ at $r \sim 0.1$ kpc. 

The gas cooling time is: $t_{\rm cool} = 1.5 k_{\rm B} T / \left( n \Lambda_{\rm cool} \right)$, 
where $\Lambda_{\rm cool}$ is the net cooling rate. 
We adopt the cooling rate from the analytical formula (Equation 9) in \citet{Fujita16}, 
which is a function of temperature and metallicity. 
It accounts for Bremsstrahlung ($\propto T^{0.5}$) and metal line cooling ($\propto T^{-1}$) terms, 
and approximates the cooling function derived by \citet{Sutherland93}. 
The $t_{\rm cool}$ radial profiles are plotted in the bottom-right panel of Fig.~\ref{fig-GasProperties_vs_R_7runs}. 
The cooling time becomes shorter than $1$ Gyr at $r < 20$ kpc. 
At the same time we note that this $t_{\rm cool}$ estimate includes Bremsstrahlung and metal line cooling only, 
and does not include radiative heating from photoionizing radiation of the UV/X-ray background. 
The $t_{\rm cool}$ in run {\it unf10kLoCA} has a gently flattening core 
with no central abrupt fall, which are observationally favorable features. 
This affirms that both cold accretion and a low efficiency $(\epsilon_f = 0.002)$ 
are crucial to produce a proper self-regulated AGN feedback loop. 




\section{Discussion} 
\label{sec-discussion} 

In this study we explore a distinct mode of AGN feedback: 
kinetic coupling of the energy output from the central BH. 
It is implemented within our larger suite of sub-resolution models: 
metal-dependent radiative cooling and heating, star-formation, stellar evolution and chemical enrichment. 
We perform simulations of isolated cool-core clusters, 
starting from an idealized set-up of a hydrostatic atmosphere with no stars. 
Our simulations produce a stellar mass of a few times $10^{10} M_{\odot}$ after $2$ Gyr; 
since forming stars with our IC is a slow process, with timescale of Gyrs, slower than timescale for feedback onset. 

We do not want to perform any comparison with observational data, 
because our initial condition do not resemble a real poor cluster, 
and hence have not done any parameter tuning here to match observations. 
Our aim is to explore the results of the new kinetic AGN feedback numerical implementation 
and assess its effectiveness, emphasising on the differences from our thermal feedback. 
The current work is also a step toward building a model of unified AGN feedback in cosmological simulations; 
which we will run in the future, and where we will do the relevant parameter tuning. 


At our current resolution, we do not resolve the physical blast of thermal feedback. 
Also at our resolved scales, 
we are not simulating the real AGN jet propagation using our kinetic feedback numerical scheme. 
Here we only attempt to capture the effect of the energy deposition 
of the mechanical jets from AGN at large ($> 20$ kpc) scales. 

Our finding that the cool core is destroyed within $\sim 5$ duty cycles using a fixed BH power of 
$\dot{E}_{\rm feed} = 10^{45}$ erg/s, is inconsistent with observational X-ray data. 
The lower BH power $\dot{E}_{\rm feed} = 2 \times 10^{44}$ erg/s 
is a more observationally consistent parameter value to maintain the cool core. 
Previous simulations \citep[e.g.,][]{Gaspari11a} have found that BHs, 
which output a fixed power with fixed duty cycles, are not properly self-regulated, 
thus they tend to produce overheating or overcooling of cluster cool cores in the long run. 

Our results reveal the challenges of devising numerical schemes for sub-resolution models, 
which must capture the physical processes given the numerical resolution. 
With regard to implementing AGN feedback in galaxy-scale simulations, 
the AGN output power is injected to the gas located near to the BH. 
The gas in these central regions can be very dense, 
hence immediately and non-physically can radiate the injected energy away. 
Such problems are numerical in nature, which especially impacts our thermal feedback results. 

We demonstrate that our kinetic AGN feedback is more effective than thermal AGN feedback, 
when implemented together with cooling, SF, stellar evolution and chemical enrichment. 
This statement might sound unphysical. 
But as \citet{Barai14} discussed in detail, the implementation of BH thermal feedback is less efficient 
within the framework of the multiphase effective sub-resolution model of star-formation by \citet{SH03}, 
where SF is based on a density threshold only. 
The thermal energy deposited to the gas particles which are multiphase (and star-forming) 
is radiated away quickly, since they are dense. 
And they attain the effective equation-of-state temperature dictated by their density. 

The feedback efficiency $\epsilon_f$ (Eq.~\ref{eq-Edot-Feed}) is an important parameter of AGN feedback models, 
which determines what fraction of the energy from the BH couples to the surrounding gas. 
Cosmological simulations generally calibrate its value to reproduce the BH mass versus 
host galaxy stellar bulge mass or velocity dispersion relations at $z = 0$ \citep[e.g.,][]{Magorrian98}. 
Various studies have used different values: 
$\epsilon_f = 0.05$ together with $\epsilon_r = 0.1$ \citep{SDH05}; 
increasing to $\epsilon_f = 0.2$ during radio-mode 
when the BH accretion rate is smaller than $0.01$ times the Eddington rate \citep{Fabjan10}; 
$\epsilon_r = 0.2$ with $\epsilon_f = 0.2$ in quasar-mode and $\epsilon_f = 0.8$ in radio-mode \citep{Ragone13}; 
$\epsilon_f = 0.15$ \citep{Booth09}; 
$\epsilon_f = 0.15$ in quasar-mode and $\epsilon_f = 1$ in radio-mode \citep{Dubois13}. 


Using such high values of $\epsilon_f$, on the other hand, 
leads to the overheating of the cool core (Fig.~\ref{fig-Evol_Gas_Star_BH}), 
which is in contrast to X-ray observations depicting a tight self-regulated balance between 
heating and cooling (e.g., preserving the observed positive X-ray temperature profile gradient for several Gyr). 
For instance, \citet{Gaspari11a, Gaspari11b, Gaspari12a, Gaspari12b} 
find, in high-resolution simulations of isolated clusters, that to quench cooling flows, 
at the same time avoiding both overheating and overcooling, 
the typical mechanical efficiency $(= \epsilon_r \epsilon_f)$ for poor clusters is $\sim 10^{-3}$. 
\citet{Prasad15} saw that AGN feedback is able to suppress cooling/SF 
using a feedback efficiency as low as $\sim 10^{-4}$, 
and the simulations show cold gas and jet cycles even after several Gyr. 
Factoring out the $\epsilon_r$, these efficiencies translate to $\epsilon_f = 10^{-3} - 0.01$, i.e., 
lower than that typically used in cosmological simulations. 
Such low $\epsilon_f$ can be analytically retrieved by equating the observed jet power to the core X-ray luminosity. 
Observationally, \citet{Merloni08} solved the continuity equation for the black hole mass function 
using the locally detected one as boundary condition, 
and the hard X-ray luminosity function of the AGN growth rate distribution, 
and found that the kinetic efficiency  is a few $10^{-3}$. 


Our choice of default $\epsilon_f = 0.02$ in this work is motivated by cosmological simulations, 
which uses similar order values, 
also we have explored $\epsilon_f = 0.002$ and $0.2$ to cover the parameter space. 


Supernovae-driven kinetic feedback is not included in our simulations, 
so that the central SMBH is the sole source of energy feedback, 
and to unambiguously infer any outflow as driven by AGN. 
In most of our second-series runs with cooling, SF and BH growth, 
the cool core is heated up by strong AGN feedback. 
The influence of supernovae feedback is expected to be sub-dominant at the cluster center. 

The more efficient kinetic feedback that we find here compared to thermal feedback 
(used in all our cluster simulations before) 
is promising in helping to solve problems pointed out by \citet{Ragone13}, 
in particular the still too large mass and too big effective radius of the brightest cluster galaxies. 


\section{Summary and Conclusion} 
\label{sec-conclusion} 

We investigate different models of AGN feedback in cluster simulations, using the SPH code {\sc GADGET-3}. 
We implement novel methods to couple the feedback energy 
from a SMBH in the {\it kinetic} form, where the velocity of the neighbouring gas is incremented. 
Gas particles lying inside a bi-conical volume around the BH 
(cones of slant height $h_{\rm BH}$ and opening angle $60^{\circ}$ along two opposite directions) 
are stochastically selected and imparted a one-time $v_w$ velocity boost. 
This renders our code able to generate outflows driven by AGN feedback in a self-consistent way. 


We perform simulations of isolated cluster of total mass $10^{14} h^{-1} M_{\odot}$, 
containing $2.1 \times 10^6$ gas particles, using a gravitational softening length of $2 h^{-1}$ kpc. 
The cluster is first evolved using hydrodynamical interactions only for $3$ Gyr, 
subsequently using cooling and hydrodynamics for further $3$ Gyr, 
which forms a dense cool core, which is taken as our initial condition. 
A collisionless BH particle is created at this time (renamed $t = 0$), 
residing at the cluster center, having a seed BH mass of $10^5 M_{\odot}$. 

Our simulations form two broad series, each having the same non-AGN sub-resolution physics, 
and the runs in it explore different AGN feedback models. 
The first series includes simulations with cooling-only, no-SF, constant energy feedback from BH at fixed duty-cycle; 
the second series includes cooling, SF, BH undergoing gas accretion, mass growth and resulting feedback. 
In particular, five of our runs in the latter series employ the ``unified'' AGN model presented 
in \citet{Steinborn15}. 
Here, we use the mechanical outflow power as the source of energy for our kinetic feedback, 
and {\it build an improved form of unified AGN feedback model}.  








Our main results are: 

\begin{itemize} 

\item 
Bipolar bubble-like outflows form, originating from the BH, and propagating 
radially outward to a distance of a few $100$ kpc. 
The cluster cool core is shock heated in several runs, to a median temperature $T > 10^{7}$ K, 
which corresponds to a central entropy flattening, $S = k_{\rm B} T / n^{2/3} > 15$ keV cm$^2$. 
A large outflow originates during each AGN-on active period. 

\item 
The output BH power $\dot{E}_{\rm feed} = 10^{45}$ erg/s heats up the cool core 
in the runs where the kicked particles are always hydrodynamically coupled. 
Wind particles decoupled from hydrodynamics 
do not have any heating effect at the center, unless the power is increased by a factor of $4$. 
BH feedback is less efficient in two runs (one with a lower BH power) where the initial cool core remains. 

\item 
When allowing the BH to grow, the BH accretion rate has a power-law increase, 
to reach a peak of few times $(1 - 10)~M_{\odot}$/yr (values depending on the feedback model). 
After the peak, the $\dot{M}_{\rm BH}$ reduces $10 - 100$ times in the runs with kinetic feedback 
and remains almost constant with thermal feedback. 

\item 
The total star formation rate increases with time initially; 
then the growing BH accretes in, heats up, and ejects out some central gas, quenching SF. 
The SFR is reduced $10$ times with thermal, and $100$ times by kinetic feedback. 
{\it Thus, kinetic feedback is more efficient in quenching the SFR.} 
The same is true for the inner cool gas mass. 

\item 
Kinetic feedback is more effective and has a stronger impact on the cluster central gas 
than thermal feedback using the same feedback efficiency, within our implementation of star-formation model. 

\item 
The inclusion of {\it cold gas accretion} in the simulations with BH growth 
shows huge variability of the $\dot{M}_{\rm BH}$ and inner cool/hot gas masses, 
thus {\it produces naturally a duty cycle of the AGN with a periodicity of $100$ Myr}. 

\item 
The unified feedback model, where we use our kinetic scheme to distribute the mechanical AGN power, 
is basically as good as the pure kinetic one in quenching the SFR, 
heating the cool core and raising its central entropy level. 
We emphasize again that the same result cannot be obtained using purely thermal feedback.

\end{itemize} 

In this paper we have presented 
an implementation of AGN feedback in the kinetic form in an SPH code. 
The results presented here further demonstrate the relevance of explicitly describing 
the mechanical nature of the radio-mode AGN feedback to create a self-regulated cooling-heating cycle 
in the core regions of isolated halos having the size of a galaxy cluster. 
We consider this as quite encouraging in view of the forthcoming application of 
our implementation of mechanical AGN feedback in cosmological simulations. 


\section*{Acknowledgments} 

We are most grateful to Volker Springel for allowing us to use the GADGET-3 code. 
We thank Lisa Steinborn, Klaus Dolag, and Alexander Beck for useful discussions. 
Simulations were run at the CINECA Supercomputing Center, 
with CPU time assigned through an ISCRA Class-C proposal, and through an INAF-CINECA grant. 
Data postprocessing and storage has been done on the CINECA facility PICO, 
granted to us thanks to our expression of interest. 
This work is supported 
by the PRIN-MIUR 201278X4FL ``Evolution of cosmic baryons'' funded by the Italian Ministry of Research, 
by the PRIN-INAF 2012 grant ``The Universe in a Box: Multi-scale Simulations of Cosmic Structures'', 
by the INDARK INFN grant, and by Consorzio per la Fisica di Trieste. 
M.G. is supported by NASA through Einstein Postdoctoral Fellowship Award Number PF-160137 
issued by the Chandra X-ray Observatory Center, which is operated by the 
Smithsonian Astrophysical Observatory for and on behalf of NASA under contract NAS8-03060.

%


\end{document}